\documentclass[aps,pra,amsmath,reprint,floatfix]{revtex4-1}
\usepackage{microtype}
\usepackage{amssymb}
\usepackage{graphicx}
\usepackage[usenames,dvipsnames]{xcolor}
\usepackage{bm}
\usepackage{times}

\renewcommand{\vec}[1]{{\bm{#1}}}
\newcommand{\md}{\mathrm{d}}
\newcommand{\mK}{{\mathrm{K}}}
\newcommand{\Ac}{\mathcal{A}}
\renewcommand{\Re}{\mathrm{Re}\!\;}
\renewcommand{\Im}{\mathrm{Im}\!\;}
\newcommand{\FG}{{\mathrm{FG}}}
\newcommand{\IFG}{{\mathrm{IFG}}}
\newcommand{\IF}{{\mathrm{IF}}}
\newcommand{\kp}{{\mathrm{kp}}}
\newcommand{\ex}{{\mathrm{ex}}}
\newcommand{\mi}{\mathrm{i}}
\newcommand{\mH}{\mathrm{H}}
\newcommand{\mI}{\mathrm{int}}
\newcommand{\mL}{\mathrm{L}}
\newcommand{\mR}{\mathrm{R}}
\newcommand{\nint}{\!\int\!}
\newcommand{\lala}{\langle\mspace{-3mu}\langle}
\newcommand{\rara}{\rangle\mspace{-3mu}\rangle}

\RequirePackage[
  hyperindex,colorlinks,bookmarksnumbered,
  plainpages=true,pdfstartview=FitH]{hyperref}
\hypersetup{urlcolor={blue!50!black}, linkcolor={red!50!black},citecolor={blue!50!black}} 
\usepackage{hyperref}
\usepackage[all]{hypcap}

\begin{document} 

\title{Improved estimator for numerical renormalization group calculations of the self-energy}
\author{Fabian B.~Kugler}
\affiliation{Department of Physics and Astronomy, Rutgers University, Piscataway, NJ 08854, USA}
\date{\today}

\begin{abstract}
We present a new estimator for the self-energy based on a combination of two equations of motion
and discuss its benefits for numerical renormalization group (NRG) calculations.
In challenging regimes, NRG results from the standard estimator, a ratio of two correlators,
often suffer from artifacts:
The imaginary part of the retarded self-energy is not properly normalized and, at low energies, overshoots to unphysical values and displays wiggles.
We show that the new estimator resolves the artifacts in these properties
as they can be determined directly from the imaginary parts of auxiliary correlators and do not involve real parts obtained by Kramers--Kronig transform.
Furthermore, we find that the new estimator yields converged results with reduced numerical effort (for a lower number of kept states) and thus is highly valuable when applying NRG to multiorbital systems.
Our analysis is targeted at NRG treatments of quantum impurity models, 
especially those arising within dynamical mean-field theory,
but most results can be straightforwardly generalized to other impurity or cluster solvers.
\end{abstract}

\maketitle

\section{Introduction}
Quantum impurity systems,
a small number of interacting degrees of freedom embedded in a noninteracting bath,
play an important role in many-body physics.
On the one hand, they are fascinating on their own right,
serving as a paradigm for strong-coupling phenomena
and as the underlying model of quantum dot devices \cite{Hanson2007}.
On the other hand, they gained much attention recently
in the study of strongly correlated lattice systems
within the dynamical mean-field theory (DMFT) \cite{Georges1996}.

For conventional quantum impurity models, 
central dynamic correlation functions are, e.g., the 
spectral function (local density of states) or the magnetic susceptibility.
By contrast, in DMFT, the quintessential object is the
(local but frequency-dependent) self-energy.
It enters many observables, such as the momentum-dependent spectral function 
(used to describe angle-resolved photoemission spectroscopy),
all types of conductivities in transport measurements,
nonlocal susceptibilities that make up structure factors,
and is needed to determine the Fermi-liquid parameters that pervade most low-energy properties.
Moreover, for almost all lattices---the popular Bethe lattice being an exception---%
the self-energy is the crucial ingredient of the DMFT self-consistency iteration \cite{Georges1996}.

The numerical renormalization group (NRG) \cite{Wilson1975} is the gold standard for solving quantum impurity models \cite{Bulla2008}.
It is often used as a real-frequency impurity solver for DMFT,
in Hubbard models with one
\cite{Bulla1999,Bulla2001,Deng2013,Lee2017,Vucicevic2019,Vranic2020,Vucicevic2021a,*Vucicevic2021b},
two \cite{Pruschke2005,Peters2010a,*Peters2010b,Peters2011,Greger2013a,Greger2013b,Kugler2021},
and three orbitals \cite{Stadler2016,Stadler2019,Kugler2019,Stadler2021},
and recently even for realistic material systems \cite{Kugler2020}. 
Modern formulations of NRG, a.k.a.\ full density-matrix (fdm) NRG
\cite{Peters2006,Weichselbaum2007},
give very accurate results for correlation functions of local operators.
Yet, the self-energy $\Sigma$ is no such correlation function 
but an irreducible vertex object,
and must be computed by different means.
Since a direct inversion of the Dyson equation is numerically disadvantageous,
$\Sigma$ is routinely computed through an equation of motion (eom)
as a quotient between two correlators \cite{Bulla1998}.
In challenging (e.g., multiorbital) situations, however, 
the results for $\Sigma$ are not always as accurate as one expects from NRG.
First, its spectral weight is not guaranteed to be properly normalized in fdm NRG,
so that the (analytically known) high-frequency asymptote may be violated.
Moreover, the imaginary part of the retarded self-energy, $\Im \Sigma_\nu$,
can overshoot to positive values at low energies 
even though causality requires $\Im \Sigma_\nu \!\leq\! 0$.
This is often accompanied by wiggles in small values of $\Im \Sigma_\nu$.

In fact, while the high-energy resolution of NRG can be increased by averaging techniques
\cite{Zitko2009,Lee2016,Lee2017}, 
these tricks do not help much in resolving the problems of $\Im\Sigma_\nu$ at low energies---%
where NRG is most powerful.
So far, the overshooting and wiggles in $\Im\Sigma_\nu$ could only be tackled by brute-force
increase of numerical effort (increasing the number of kept states), 
so that accurate results for $\Sigma$ were a computational bottleneck.

In this paper, we present a new formula for the self-energy,
based on a combination of a one- \cite{Bulla1998} and twofold \cite{Kaufmann2019} application of the eom.
This result strongly alleviates the previously mentioned artifacts:
The high-frequency asymptote of $\Re \Sigma$ is fulfilled exactly,
overshooting of $\Im \Sigma$ is ruled out, and
the value of $\Im \Sigma$ at zero energy is improved by several orders of magnitude.
Our formula involves three instead of two \cite{Bulla1998} correlators.
While this naively increases the numerical costs by a factor of $1.5$,
we find that accurate results with the new formula
are obtained already with less numerical effort
(a lower number of kept states) compared to the standard scheme.
Hence, our approach also makes NRG computations of $\Sigma$ more efficient 
and thus helps to equip DMFT+NRG with the tools needed for treating Hubbard models with ever more orbitals.

The rest of the paper is organized as follows.
In Sec.~\ref{sec:overview}, we give an overview of the theoretical framework as well as the previous and new self-energy estimators.
The derivation of these expressions and their properties is found in the subsequent Sec.~\ref{sec:derivation}.
In Sec.~\ref{sec:numerics}, we demonstrate the benefits of the new approach with numerical results.
There, we start with the single-orbital Anderson impurity model and proceed with one-, two-, and three-orbital Hubbard models treated in DMFT.
Section~\ref{sec:conclusion} contains our conclusions,
Appendix~\ref{sec:appendix1} discusses the generalization to matrix-valued correlation functions,
and Appendix~\ref{sec:appendix2} provides additional numerical data.

\section{Overview}
\label{sec:overview}
\subsection{Definitions}
Quantum impurity models are naturally divided into the interacting impurity and the noninteracting bath.
We denote electron creation operators of the former by $d^\dag_\alpha$ and those of the latter by $c^\dag_{k\alpha}$.
The index $\alpha$ enumerates spin ($\sigma$) and possibly orbital ($m$) quantum numbers;
the bath modes are further labeled by $k$, standing, e.g., for momentum.
The noninteracting part of the Hamiltonian $H \!=\! H_0 + H_\mI$ generally reads
\begin{align}
H_0
& =
\textstyle
\sum_\alpha
\epsilon_{d,\alpha} d^\dag_\alpha d_\alpha
+
\sum_{k,\alpha}
\epsilon_{k\alpha} c^\dag_{k\alpha} c_{k\alpha}
\nonumber
\\
& \ 
\textstyle
+
\sum_{k,\alpha} ( V_{k\alpha} d_\alpha^\dag c_{k\alpha} + \mathrm{H.c.} )
.
\label{eq:H0}
\end{align}
For the interacting part, we consider two examples.
The single-orbital ($\alpha \!=\! \sigma$) Anderson impurity model 
\cite{Anderson1961} has
\begin{align}
H_\mI
=
U n_\uparrow n_\downarrow
, 
\quad
n_\sigma = d^\dag_\sigma d_\sigma
.
\label{eq:Anderson}
\end{align}
In the multiorbital case [$\alpha=(\sigma,m)$],
we use the generalization of Eq.~\eqref{eq:Anderson}
introduced by Dworin and Narath \cite{Dworin1970,Georges2013},
\begin{align}
H_\mI
=
\tfrac{3}{4} J N
+ \tfrac{1}{2} (U - \tfrac{3}{2} J) N (N-1) 
- J \vec{S}^2
,
\label{eq:Dworin}
\end{align}
where $N = \sum_\alpha d^\dag_\alpha d_\alpha$
and $\vec{S} = \sum_{\sigma \sigma' m} d^\dag_{\sigma' m} \vec{\tau}_{\sigma'\sigma} d_{\sigma m}$
with the Pauli matrices $\vec{\tau}$.

We will be interested in correlation functions involving the fundamental operators $d_\alpha$, $d_\alpha^\dag$ as well as the auxiliary operators
\begin{align}
\label{eq:def_q}
q_\alpha 
=
[d_\alpha, H_\mI]
,
\quad
q_\alpha^\dag
=
[H_\mI, d_\alpha^\dag]
.
\end{align}
They allow us to define four fermionic correlation functions:
\begin{subequations}
\label{eq:def_G_I_F}
\begin{flalign}
\qquad
G_{\alpha z}  
& = 
\lala d_\alpha, d_\alpha^\dag \rara_z
,
&
\quad
I_{\alpha z} 
& = 
\lala q_\alpha, q_\alpha^\dag \rara_z
,
&
\\
\qquad
F^\mL_{\alpha z} 
& =
\lala q_\alpha, d_\alpha^\dag \rara_z
,
&
\quad
F^\mR_{\alpha z} 
& =
\lala d_\alpha, q_\alpha^\dag \rara_z
.
&
\end{flalign}
\end{subequations}
Here, our notation follows Ref.~\cite{Bulla1998}:
$z$ is a complex frequency variable.
It can be a discrete imaginary frequency, $\mi\nu$, or a continuous real frequency $\nu$.
In the former case, $\lala A, B \rara_z$ is the Fourier transform of the imaginary-time correlator
$- \langle \mathit{T} A(\tau) B \rangle$,
with the time-ordering operator $\mathit{T}$.
In the latter, it corresponds to the retarded correlator
$- \mi\theta(t) \langle \{ A(t), B \} \rangle$,
with the step function $\theta$ and the anticommutator $\{ \cdot , \cdot \}$.

In systems defined by Eqs.~\eqref{eq:H0}--\eqref{eq:Dworin}, the fermionic correlation functions are diagonal in $\alpha$ 
(and thus carry only a single subscript). It then follows (as shown below) that $F^\mL_{\alpha z} = F^\mR_{\alpha z}$;
we will hence mostly drop the superscript.
Appendix~\ref{sec:appendix1} addresses the case where $H_0$ has off-diagonal contributions
and the correlation functions become matrix-valued.
Then, $F^\mL$ and $F^\mR$ are not equal, but still related by symmetry.
For a close connection of both situations, we often use matrix-type notation in the main text, too,
and restore the superscripts $\mL$, $\mR$ in key places.
Moreover, even for $\alpha$-diagonal computations, $F^\mL_{\alpha z} = F^\mR_{\alpha z}$
might be slightly violated numerically. 
It may then be helpful to use the matrix-type formulas, which are symmetric in $F^\mL$ and $F^\mR$.

Before moving on to the self-energy, let us briefly recall
how correlators like $G_{\alpha z}$, $F_{\alpha z}$, and $I_{\alpha z}$ 
are obtained in NRG.

\subsection{NRG correlation functions}
In NRG, a general correlator $C_{\alpha z}$
is first computed as a discrete version of the spectral part $-\tfrac{1}{\pi} \Im C_{\alpha \nu}$. 
After broadening $\Im C_{\alpha \nu}$,
the real part follows by Kramers--Kronig transform as
\begin{align}
\Re C_{\alpha \nu}
=
- 
\frac{1}{\pi}
\, \mathcal{P} \! \nint \md \nu' \,
\frac{\Im C_{\alpha \nu'}}{\nu - \nu'}
.
\label{eq:KramersKornig}
\end{align}
By construction of fdm NRG, 
the total weight of $\Im C_{\alpha \nu}$ is guaranteed to be exact \cite{Peters2006,Weichselbaum2007}.
Further, by the very nature of NRG, results for $\Im C_{\alpha \nu}$ are most accurate at low energies.
Going to larger frequencies, $\Im C_{\alpha \nu}$ can be significantly less accurate,
reflecting the logarithmic discretization of the hybridization function.
Refined averaging and adaptive broadening techniques \cite{Zitko2009,Lee2016,Lee2017}
help to minimize overbroadening.
Yet, the approximate nature of $\Im C_{\alpha \nu}$ for large $\nu$ remains.
In particular, it is known that the moments
$C^{(n)} = - \tfrac{1}{\pi} \nint \md \nu \, \nu^n \Im C_{\alpha \nu}$ are \textit{not} reproduced exactly for $n>0$.
Now, by Eq.~\eqref{eq:KramersKornig}, the large-energy inaccuracies of $\Im C_{\alpha \nu}$ 
are not only passed down to $\Re C_{\alpha \nu}$ but are also spread in frequency space.
Hence, in the following, we will aim to minimize the effect of real parts of correlation functions in the computation of $\Sigma$.

\subsection{Self-energy formulas}
The self-energy is defined by the Dyson equation as
\begin{align}
\Sigma_{\alpha z} 
= 
(G^0_{\alpha z})^{-1} - (G_{\alpha z})^{-1}
.
\label{eq:Dyson}
\end{align}
Here, $G^0_{\alpha z}$ is the bare propagator,
which can be written in terms of the hybridization function
$\Delta_{\alpha z}$ as
\begin{flalign}
\label{eq:G0_Delta}
(G^0_{\alpha z})^{-1}
& =
z - \epsilon_{d,\alpha} - \Delta_{\alpha z}
,
\qquad
\Delta_{\alpha z} 
= 
\sum_k \frac{|V_{k\alpha}|^2}{ z-\epsilon_{k\alpha} }
.
\hspace{-0.5cm}
&
\end{flalign}
The retarded self-energy fulfills the Kramers--Kronig relation
\begin{align}
\Re \Sigma_{\alpha \nu}
=
\Sigma^\mH_\alpha
- 
\frac{1}{\pi}
\, \mathcal{P} \! \nint \md \nu' \,
\frac{\Im \Sigma_{\alpha \nu'}}{\nu - \nu'}
,
\label{eq:KramersKornig_SE}
\end{align}
where $\Sigma^\mH_\alpha$ is the constant Hartree part.
This relates the high-frequency asymptote of the real part to the total weight in the imaginary part.
We define the $1/\nu$ coefficient of $\Re \Sigma_{\alpha \nu}$
(or the $1/z$ coefficient of $\Sigma_{\alpha z}$)
as the moment $\Sigma_\alpha^{(0)}$ which fulfills
\begin{align}
\Sigma_\alpha^{(0)}
= 
\lim_{|\nu|\to\infty} \nu \, (\Re\Sigma_{\alpha \nu}-\Sigma^\mH_\alpha) 
= 
- \frac{1}{\pi}
\nint \md \nu \, \Im \Sigma_{\alpha \nu}
.
\label{eq:SigmaReIm}
\end{align}

For many algorithms that yield $G_{\alpha z}$ directly, Eq.~\eqref{eq:Dyson} is not ideal to extract $\Sigma$. In NRG, it is basically inapplicable since $G^0_{\alpha z}$ involves the exact, continuous hybridization function. By contrast, $G_{\alpha z}$ is the result of an approximate calculation where $\Delta_{\alpha z}$ was discretized. Thus, cancellations between $G^0_{\alpha z}$ and $G_{\alpha z}$ required for Eq.~\eqref{eq:Dyson} do not work properly and induce large numerical errors.
For this reason, NRG self-energies are routinely computed by means of an eom yielding \cite{Bulla1998}
\begin{align}
\Sigma_{\alpha z}^\FG
=
F_{\alpha z} (G_{\alpha z})^{-1}
.
\label{eq:SigmaFG_1}
\end{align}
Our new formula, based on a combination of a one- and twofold \cite{Kaufmann2019} application of the eom, reads
\begin{align}
\Sigma_{\alpha z}^\IFG
=
\Sigma^\mH_\alpha + I_{\alpha z} - F^\mL_{\alpha z} (G_{\alpha z})^{-1} F^\mR_{\alpha z}
.
\label{eq:SigmaIFG_1}
\end{align}
We restored the superscript on $F$ in light of matrix-valued applications (see Appendix~\ref{sec:appendix1}). In the given $\alpha$-diagonal setting, the last term can be simply written as $F_{\alpha z}^2/G_{\alpha z}$.
Now, what are the advantages of Eq.~\eqref{eq:SigmaIFG_1} over Eq.~\eqref{eq:SigmaFG_1}?

Focusing on the imaginary part,
from Eq.~\eqref{eq:SigmaFG_1}, we get
\begin{align}
\Im \Sigma^\FG_{\alpha \nu}
& =
\frac{ \Im F_{\alpha \nu} \Re G_{\alpha \nu}
- \Re F_{\alpha \nu} \Im G_{\alpha \nu} }
{ |G_{\alpha \nu}|^2 }
.
\label{eq:ImSigmaFG}
\end{align}
Evidently, $\Im \Sigma^\FG_{\alpha \nu}$ is determined by the imaginary parts \textit{and} the real parts of NRG correlators. 
One finds that the total weight $\nint \md \nu\, \Im \Sigma^\FG_{\alpha \nu}$ typically does not give the exact value.
Further, due to the real parts involved, $\Im \Sigma^\FG_{\alpha \nu}$ at low energies is less accurate than one is used to for imaginary parts of correlators computed \textit{directly} with NRG.
In challenging regimes, one encounters the aforementioned artifacts that $\Im \Sigma^\FG_{\alpha \nu}$ overshoots to positive values and displays wiggles for low $\nu$.

By contrast, for $\Sigma_{\alpha z}^\IFG$, we will show that
both the total weight of $\Im \Sigma^\IFG_{\alpha \nu}$,
as an important high-energy property,
as well as the low-energy behavior of $\Im \Sigma^\IFG_{\alpha \nu}$
is determined by the imaginary parts of NRG correlators only.
Indeed, we have
\begin{align}
\nint \md \nu \, \Im \Sigma_{\alpha \nu}^\IFG
& =
\nint \md \nu \, \Im I_{\alpha \nu}
-
\frac{
(\nint \md \nu \, \Im F_{\alpha \nu})^2
}{
\nint \md \nu \, \Im G_{\alpha \nu}
}
,
\label{eq:SigmaIFG_weight}
\end{align}
and, for a Fermi liquid,
\begin{flalign}
\Im \Sigma^\IFG_{\alpha \nu}
& =
\Im I_{\alpha \nu}
-
\frac{ 
(\Im F_{\alpha \nu})^2
}
{ \Im G_{\alpha \nu} }
+
\kappa\mathit{O}(\nu^4,T^4,\nu^2 T^2) 
,
\hspace{-0.5cm}
&
\label{eq:SigmaIFG_lowenergy}
\end{flalign}
where $\kappa \sim 1/(T_\mK^2 \Im \Delta_{\alpha,\nu=0})$.
(A similar relation also holds for non-Fermi liquids whenever $\Im \Sigma$ is small, but the remainder term may not be as easy to estimate.)
For these imaginary parts of NRG correlators
($\Im I_{\alpha \nu}$, $\Im F_{\alpha \nu}$, $\Im G_{\alpha \nu}$),
the exact total weight in Eq.~\eqref{eq:SigmaIFG_weight} is guaranteed
and the low-energy behavior in Eq.~\eqref{eq:SigmaIFG_lowenergy} is extremely accurate.
Hence, because of Eqs.~\eqref{eq:SigmaIFG_weight} and \eqref{eq:SigmaIFG_lowenergy},
we can expect $\Sigma_{\alpha z}^\IFG$ to give better results in NRG
than $\Sigma_{\alpha z}^\FG$.
Below, we will first derive these properties analytically and then demonstrate their benefits numerically.

\section{Derivations}
\label{sec:derivation}
\subsection{Equations of motion}
\label{sec:eoms}
The starting point is the well-known equation of motion
\begin{subequations}
\label{eq:basic_eom}
\begin{align}
\langle \{ A, B \} \rangle
& =
\lala z A - [A, H], B \rara_z
,
\label{eq:basic_eom_L}
\\
\langle \{ A, B \} \rangle
& =
\lala A, z B - [H, B] \rara_z
,
\label{eq:basic_eom_R}
\end{align}
\end{subequations}
as used, e.g., in Refs.~\cite{Bulla1998,Hafermann2012,Moutenet2018,Kaufmann2019}.
In short, Eqs.~\eqref{eq:basic_eom_L} and \eqref{eq:basic_eom_R}
follow by differentiating the time-dependent two-point correlator w.r.t.\
the first and second time argument, respectively.
Then, the equal-time anticommutator stems from the time derivative of the 
(time-ordering) step function,
$z$ from the time derivative itself after Fourier transform, and
the commutator with $H$ from the the Heisenberg time evolution. 

Commutators between the bare Hamiltonian $H_0$ and the basic operators $d_\alpha$ and $d_\alpha^\dag$ can be immediately deduced as
\begin{subequations}
\begin{align}
[d_\alpha, H_0] 
& =
\textstyle
\epsilon_{d,\alpha} d_\alpha + \sum_k V_{k\alpha} c_{k\alpha}
,
\\
[H_0, d_\alpha^\dag] 
& =
\textstyle
d_\alpha^\dag \epsilon_{d,\alpha} +\sum_k c_{k\alpha}^\dag V^*_{k\alpha} 
.
\end{align}
\end{subequations}
The last summands involve bath operators.
It can easily be shown via Eqs.~\eqref{eq:basic_eom}
that, for general impurity operators $O_d$,
\begin{subequations}
\begin{align}
&
\textstyle
\sum_k V_{k\alpha} \lala c_{k\alpha}, O_d \rara_z
=
\Delta_{\alpha z} \lala d_{\alpha}, O_d \rara_z
,
\\
&
\textstyle
\sum_k \lala O_d, c^\dag_{k\alpha} \rara_z V^*_{k\alpha} 
=
\lala O_d, d_{\alpha}^\dag \rara_z \Delta_{\alpha z}
.
\end{align}
\end{subequations}

In Eqs.~\eqref{eq:basic_eom}, the equal-time term  is trivial for 
the creation and annihilation operators,
$\{d_\alpha, d_\alpha^\dag\} = 1$.
We thus get
\begin{align}
1
& =
\lala z d_\alpha - [d_\alpha, H_0], d_\alpha^\dag \rara_z
-
\lala [d_\alpha, H_\mI], d_\alpha^\dag \rara_z
\nonumber
\\
& =
( z - \epsilon_{d,\alpha} - \Delta_{\alpha z} )
\lala d_\alpha, d_\alpha^\dag \rara_z
-
\lala q_\alpha, d_\alpha^\dag \rara_z
\nonumber
\\
& =
( G^0_{\alpha z} )^{-1} G_{\alpha z} - F^\mL_{\alpha z}
.
\label{eq:1st-order_eom}
\end{align}
Using Eq.~\eqref{eq:basic_eom_R} instead of \eqref{eq:basic_eom_L} yields 
$1 \!=\! G_{\alpha z} (G^0_{\alpha z})^{-1} - F^\mR_{\alpha z}$.
In the given $\alpha$-diagonal setting, this implies 
$F^\mR_{\alpha z} \!=\! F^\mL_{\alpha z}$.

We next employ Eq.~\eqref{eq:basic_eom_R} for $F^\mL_{\alpha z}$.
This way, the commutator acts on $d_\alpha^\dag$, similarly as before.
The equal-time term with one $q_\alpha$ operator gives the Hartree self-energy, 
\begin{align}
\Sigma^\mH_\alpha 
= \langle \{ [d_\alpha, H_\mI], d_\alpha^\dag \} \rangle 
= \langle \{ d_\alpha, [H_\mI, d_\alpha^\dag] \} \rangle
.
\label{eq:SigmaH}
\end{align}
In total, we get
\begin{align}
\Sigma^\mH_\alpha
& =
\lala q_\alpha, z d_\alpha^\dag - [H_0, d_\alpha^\dag] \rara_z
-
\lala q_\alpha, [H_\mI, d_\alpha^\dag] \rara_z
\nonumber
\\
& =
\lala q_\alpha, d_\alpha^\dag \rara_z
( z - \epsilon_{d,\alpha} - \Delta_{\alpha z} )
-
\lala q_\alpha, q_\alpha^\dag \rara_z
\nonumber
\\
& =
F^\mL_{\alpha z} ( G^0_{\alpha z} )^{-1} - I_{\alpha z}
.
\label{eq:2nd-order_eom}
\end{align}
Applying Eq.~\eqref{eq:basic_eom_L} to $F^\mR_{\alpha z}$ yields
$\Sigma^\mH_\alpha \!=\! ( G^0_{\alpha z} )^{-1} F^\mR_{\alpha z} - I_{\alpha z}$.
Again, this shows $F^\mR_{\alpha z} \!=\! F^\mL_{\alpha z}$ in the $\alpha$-diagonal setting.
We will hence drop the superscript in most of the following.

\subsection{Self-energy estimators}
\label{sec:estimators}
Using the Dyson equation~\eqref{eq:Dyson},
the first-order eom result for $\Sigma_{\alpha z}$ directly follows from Eq.~\eqref{eq:1st-order_eom} as
\begin{align}
\Sigma_{\alpha z} 
& =
F_{\alpha z} G_{\alpha z}^{-1}
\equiv
\Sigma^\FG_{\alpha z} 
.
\label{eq:SigmaFG}
\end{align}
This is the famous result from Ref.~\cite{Bulla1998}.
Here and below, the expression after the $\equiv$ sign serves for future reference.
Next, the second-order formula for $\Sigma$ is obtained by inserting
the eom~\eqref{eq:2nd-order_eom} for $F_{\alpha z}$
into the first-order result~\eqref{eq:SigmaFG} for $\Sigma$:
\begin{align}
\label{eq:SigmaIG}
\Sigma_{\alpha z}
& =
( \Sigma^\mH_\alpha + I_{\alpha z} )
G^0_{\alpha z} G_{\alpha z}^{-1}
\equiv
\Sigma^{\mathrm{IG}}_{\alpha z}
.
\end{align}
Using Eq.~\eqref{eq:Dyson} for $(G_{\alpha z})^{-1}$
and isolating $\Sigma_{\alpha z}$, we get
\begin{align}
\label{eq:SigmaI}
\Sigma_{\alpha z}
& =
[1 + ( \Sigma^\mH_\alpha + I_{\alpha z} ) G^0_{\alpha z} ]^{-1}
(\Sigma^\mH_\alpha + I_{\alpha z})
\equiv
\Sigma^{\mathrm{I}}_{\alpha z}
,
\end{align}
the ``symmetric improved estimator'' derived in Ref.~\cite{Kaufmann2019}
\footnote{We note that $\Sigma^{\mathrm{I}}_{\alpha z}$ requires only one correlator, $I_{\alpha z}$,
instead of the two needed for $\Sigma^\FG_{\alpha z}$.
Yet, with the same trick that led from Eq.~\eqref{eq:SigmaIG} to \eqref{eq:SigmaI},
we can transform Eq.~\eqref{eq:SigmaFG} to
$\Sigma^{\mathrm{F}}_{\alpha z} = (1+F_{\alpha z})^{-1} F_{\alpha z} (G^0_{\alpha z})^{-1}$.
This result, too, involves only a single full correlator.
However, we numerically found $\Sigma^{\mathrm{F}}_{\alpha z}$ to be less accurate than 
$\Sigma^{\mathrm{I}}_{\alpha z}$
and hence do not discuss it any further.}.

Using Eq.~\eqref{eq:Dyson} for $(G^0_{\alpha z})^{-1}$ instead of $(G_{\alpha z})^{-1}$
in Eq.~\eqref{eq:SigmaIG}, after bringing both propagators to the left of Eq.~\eqref{eq:SigmaIG}, yields
\begin{align}
\Sigma_{\alpha z}
& =
\Sigma^\mH_\alpha + I_{\alpha z} 
-
\Sigma_{\alpha z} G_{\alpha z} \Sigma_{\alpha z}
.
\label{eq:SigmaI_rec}
\end{align}
This formula was used in Ref.~\cite{Moutenet2018} for a recursive diagrammatic Monte Carlo scheme.
Here, we process this result further by inserting the standard estimator for $\Sigma$ on the right,
$\Sigma_{\alpha z} G_{\alpha z} \Sigma_{\alpha z} \!=\! F_{\alpha z}^2 / G_{\alpha z}$,
to obtain an improved estimator on the left.
Restoring superscripts yields the symmetric expression
\begin{align}
\Sigma_{\alpha z} 
= 
\Sigma^\mH_\alpha + I_{\alpha z}
-
F^\mL_{\alpha z} (G_{\alpha z})^{-1} F^\mR_{\alpha z}
\equiv
\Sigma^\IFG_{\alpha z} 
.
\label{eq:SigmaIFG}
\end{align}
This is our main result,
as anticipated in Eq.~\eqref{eq:SigmaIFG_1},
for a new, improved estimator for NRG calculations of the self-energy.

Equation~\eqref{eq:SigmaIFG} can also derived in a different way.
First, we rephrase Eq.~\eqref{eq:1st-order_eom} as 
$(G^{\mathrm{impr}}_{\alpha z})^{-1} \!=\! ( 1 + F^\mL_{\alpha z} )^{-1} (G^0_{\alpha z})^{-1} $
and Eq.~\eqref{eq:2nd-order_eom} as 
$F^{\mR,\mathrm{impr}}_{\alpha z} \!=\! G^0_{\alpha z} ( \Sigma^\mH_\alpha + I_{\alpha z} )$.
As indicated by the superscript, we view these expressions for $G$ and $F$
as improved estimators in terms of the higher-order correlators $F$ and $I$, respectively.
Thereby, we aim for an improved $\Sigma$ estimator by means of
Eq.~\eqref{eq:SigmaFG} in the form
$\Sigma^{\mathrm{impr}}_{\alpha z} = (G^{\mathrm{impr}}_{\alpha z})^{-1} F^{\mR,\mathrm{impr}}_{\alpha z}$.
This way, $G^0_{\alpha z}$ conveniently cancels.
The expression we get is
\begin{align}
\label{eq:SigmaIF}
\Sigma_{\alpha z} 
& = 
(1 + F_{\alpha z})^{-1}
(\Sigma^\mH_\alpha + I_{\alpha z})
\equiv
\Sigma^\IF_{\alpha z} 
.
\end{align}
Yet, the denominator turns out to be numerically disadvantageous.
We thus multiply Eq.~\eqref{eq:SigmaIF} by $1 \!+\! F_{\alpha z}$ and use Eq.~\eqref{eq:SigmaFG} again in the form
$F_{\alpha z} \Sigma_{\alpha z} = F_{\alpha z}^2/G_{\alpha z}$ to reproduce Eq.~\eqref{eq:SigmaIFG}.

With $\Sigma^\FG$, $\Sigma^{\mathrm{IG}}$, $\Sigma^{\mathrm{I}}$, $\Sigma^\IFG$, $\Sigma^\IF$,
we have a total of five self-energy estimators available.
However, $\Sigma^\mathrm{IG}$ and $\Sigma^\mathrm{I}$
are not ideal for NRG since they mix full and bare correlators.
Thereby, they mix objects like $G_{\alpha z}$ and $I_{\alpha z}$, which are computed after discretization,
with the exact, continuum object $G^0_{\alpha z}$.
This hinders cancellations and often entails numerical artifacts.
As already mentioned, the denominator in Eq.~\eqref{eq:SigmaIF} 
makes $\Sigma^\IF$ numerically disadvantageous;
we will elaborate on this in Sec.~\ref{sec:denominator}.
Consequently, $\Sigma^\mathrm{FG}_{\alpha z}$ and $\Sigma^\mathrm{IFG}_{\alpha z}$
are the most suitable estimators for NRG. Next, we derive the properties of
their high- and low-energy behavior anticipated before.

\subsection{High- and low-energy behavior}
We start with the high-energy behavior.
In Eq.~\eqref{eq:SigmaReIm}, we defined the self-energy moment $\Sigma_\alpha^{(0)}$,
which represents the total weight of $\Im \Sigma_{\alpha \nu}$ as well as the first term
in a high-frequency expansion of $\Re \Sigma_{\alpha \nu}$.
Via the second property, Eq.~\eqref{eq:SigmaIFG_weight} can be derived in a few steps.

Let us consider again a general correlator $C_z = \lala A, B \rara_z$ 
with $C_{|z|\to\infty} = 0$,
as a placeholder for $G_{\alpha z}$, $F_{\alpha z}$, and $I_{\alpha z}$.
The spectral representation implies the high-frequency expansion
\begin{align}
C_z 
& = 
\sum_{n=1}^\infty \frac{ C^{(n-1)} }{ z^n }
,
\quad
C^{(n)}
=
-
\frac{1}{\pi}
\nint \md \nu \, \nu^{n} \Im C_{\alpha \nu}
.
\label{eq:high-frequency_expansion}
\end{align}
The $C^{(n)}$ can also be obtained from expectation values, as
\begin{align*}
C^{(0)}
=
\langle \{ A, B \} \rangle
, \quad
C^{(1)}
=
\langle \{ [A, H], B \} \rangle
=
\langle \{ A, [H, B] \} \rangle
,
\end{align*}
etc. The leading coefficients for our specific correlators are
$G^{(0)}_\alpha = 1$, $F^{(0)}_\alpha = \Sigma^\mH_\alpha$, 
$I^{(0)}_\alpha = \langle \{ q_\alpha, q_\alpha^\dag \} \rangle$. 
For the self-energy estimators, we can then easily deduce
\begin{subequations}
\begin{align}
\Sigma^\FG_{\alpha z}
& =
\frac{ F^{(0)}_\alpha }{ G^{(0)}_\alpha }
+
\Bigg[
\frac{ F^{(1)}_\alpha }{ G^{(0)}_\alpha }
-
\frac{ F^{(0)}_\alpha G^{(1)}_\alpha }{ \big(G^{(0)}_\alpha\big)^2 }
\Bigg]
\frac{1}{z}
+ \mathit{O}\bigg( \frac{1}{z^2} \bigg)
,
\label{eq:high-freq_SigmaFG}
\\
\Sigma^\IFG_{\alpha z}
& =
\Sigma^\mH_\alpha
+
\Bigg[
I^{(0)}_\alpha
-
\frac{ \big( F^{(0)}_\alpha \big)^2 }{ G^{(0)}_\alpha }
\Bigg]
\frac{1}{z}
+ \mathit{O}\bigg( \frac{1}{z^2} \bigg)
.
\label{eq:high-freq_SigmaIFG}
\end{align}
\end{subequations}
The combination of  Eqs.~\eqref{eq:SigmaReIm}, \eqref{eq:high-frequency_expansion}, and \eqref{eq:high-freq_SigmaIFG} implies Eq.~\eqref{eq:SigmaIFG_weight}.

As mentioned before,
the exact $C^{(0)}$ is guaranteed by the sum-rule conserving fdm NRG \cite{Peters2006,Weichselbaum2007}.
However, $C^{(1)}$ is much less accurate as it
probes $\Im C_\nu$ with increasing weight at large $\nu$
and thus suffers from NRG discretization artifacts.
With the standard estimator $\Sigma^\FG_{\alpha z}$,
the exact coefficients $F^{(0)}_\alpha$ and $G^{(0)}_\alpha$ generate the exact Hartree term 
$\Sigma^\mH_\alpha = F^{(0)}_\alpha / G^{(0)}_\alpha$.
Yet, $\Sigma^\mH_\alpha$ is also readily available via expectation values,
see Eq.~\eqref{eq:SigmaH}, 
whereas the moment $\Sigma_\alpha^{(0)}$ 
in $\Sigma^\FG_{\alpha z}$
involves coefficients $F^{(1)}_\alpha$ and $G^{(1)}_\alpha$
and is thus not very accurate.
By contrast, $\Sigma^\IFG_{\alpha z}$ takes $\Sigma^\mH_\alpha$ as input
and uses the exact coefficients $I^{(0)}_\alpha$, $F^{(0)}_\alpha$, $G^{(0)}_\alpha$
to generate the exact self-energy moment 
$\Sigma_\alpha^{(0)}$.

Next, we take a closer look at $\Im \Sigma$ at low energies.
For $\Sigma^\FG_{\alpha z}$, Eq.~\eqref{eq:ImSigmaFG} directly follows from
Eq.~\eqref{eq:SigmaFG_1} and requires no further comment.
Deriving Eq.~\eqref{eq:SigmaIFG_lowenergy} for $\Sigma^\IFG_{\alpha z}$
takes only two steps.
Straightforward algebra yields
\begin{flalign}
\Im \Sigma^\IFG_{\alpha \nu}
& =
\Im I_{\alpha \nu}
-
\Im
\bigg(
\frac{F_{\alpha \nu}^2}{G_{\alpha \nu}}
\bigg)
&
\nonumber
\\
& =
\Im I_{\alpha \nu}
-
\frac{ (\Im F_{\alpha \nu})^2 }
{ \Im G_{\alpha \nu} }
+
\frac{ |G_{\alpha \nu}|^2 }
{ \Im G_{\alpha \nu} }
( \Im \Sigma^\FG_{\alpha \nu} )^2
.
\hspace{-0.5cm}
&
\label{eq:ImSigmaIFG_derivation}
\end{flalign}
The last term, expressed through 
$\Im \Sigma^\FG_{\alpha \nu}$ of Eq.~\eqref{eq:ImSigmaFG},
is typically very small, since $\Im \Sigma_{\alpha \nu}$ is small at low frequencies.
Indeed, in a Fermi liquid,
$-\Im \Sigma_{\alpha \nu} = \mathit{O}(\nu^2 / T_\mK,T^2 / T_\mK)$
in terms of the Kondo temperature $T_\mK$,
and, furthermore, $\Im G_{\alpha \nu} / |G_{\alpha \nu}|^2 = - \Im 1/G_{\alpha \nu}$,
which gives $\Im \Delta_{\alpha, \nu=0}$ at $\nu, T \to 0$.
Using this result in Eq.~\eqref{eq:ImSigmaIFG_derivation}
yields Eq.~\eqref{eq:SigmaIFG_lowenergy}.

Equation~\eqref{eq:ImSigmaIFG_derivation} reveals an intimate connection between 
$\Im \Sigma^\IFG_{\alpha \nu}$ and $\Im \Sigma^\FG_{\alpha \nu}$.
We can infer that,
if $\Im \Sigma^\FG_{\alpha \nu}$ shows artifacts at values of $|\Im \Sigma^\FG_{\alpha \nu}|=y$ (e.g., $y \approx 10^{-3}$ in appropriate units),
then $\Im \Sigma^\IFG_{\alpha \nu}$ will show similar artifacts at values $\sim\! y^2$ (i.e., $10^{-6}$ in the example).
This quadratic relation evidently enables a huge improvement,
but it still hinders $\Im \Sigma^\IFG_{\alpha \nu}$ from reaching down all the way to zero 
in a $T \!=\! 0$ Fermi liquid.
Accordingly, for determining the Fermi-liquid parameter $\Im \Sigma_{\alpha, \nu=0}$,
it may be preferential to directly use Eq.~\eqref{eq:SigmaIFG_lowenergy},
i.e., incorporate the knowledge of Eq.~\eqref{eq:ImSigmaIFG_derivation}
where $(\Im \Sigma^\FG_{\alpha, \nu \to 0})^2$ is negligible
\footnote{
One can also use the numerical result obtained for $\Im \Sigma^\IFG_{\alpha \nu}$
and substitute it on the right of Eq.~\eqref{eq:ImSigmaIFG_derivation} \textit{instead of}
$\Im \Sigma^\FG_{\alpha \nu}$. This yields a notable improvement at low energies
but spoils high-energy properties such as the normalization of $\Im \Sigma_{\alpha \nu}$.
Using Eq.~\eqref{eq:ImSigmaIFG_derivation} with 
$\Im \Sigma^\FG_{\alpha \nu} \to \Im \Sigma^\IFG_{\alpha \nu}$ on the right 
and solving the quadratic equation for $\Im \Sigma^\IFG_{\alpha \nu}$
did not turn out to be helpful.}.

\subsection{Shifting quadratic parts in the Hamiltonian}
\label{sec:shift}
The derivations in Secs.~\ref{sec:eoms} and \ref{sec:estimators} 
build on the separation $H \!=\! H_0 + H_\mI$.
While $H_0$ is the quadratic part, Eq.~\eqref{eq:H0},
it is not specified whether or not $H_\mI$ also contains a term quadratic in $d_\alpha^{(\dag)}$.
Indeed, we may shift both $H_0$ and $H_\mI$ to 
\begin{align}
\tilde{H}_0 = H_0 + d_\alpha^\dag \zeta_\alpha d_\alpha
, \quad
\tilde{H}_\mI = H_\mI - d_\alpha^\dag \zeta_\alpha d_\alpha
.
\end{align} 
This leaves $H$ invariant; hence, it does not change any properties of the system, and all above arguments still hold.
The self-energies obtained in either way are related as
\begin{flalign}
(\tilde{G}^0_{\alpha z})^{-1} - \tilde{\Sigma}_{\alpha z} 
& = 
(G^0_{\alpha z})^{-1} - \Sigma_{\alpha z}
\ \ \Rightarrow \ \
\Sigma_{\alpha z} 
= 
\zeta_\alpha + \tilde{\Sigma}_{\alpha z}
.
\hspace{-0.5cm}
&
\end{flalign}

How does this shift affect the \textit{numerical} results for the two $\Sigma$ estimators
$\Sigma^\FG$ and $\Sigma^\IFG$?
From
$\tilde{q}_\alpha = q_\alpha - \zeta_\alpha d_\alpha$,
$\tilde{q}^\dag_\alpha = q_\alpha^\dag - d_\alpha^\dag \zeta_\alpha$,
we can directly infer that $\tilde{\Sigma}^\mH_\alpha = \Sigma^\mH_\alpha - \zeta_\alpha$ and
$\tilde{F}^\mL_{\alpha z} = F_{\alpha z} - \zeta_\alpha G_{\alpha z}$,
$\tilde{F}^\mR_{\alpha z} = F_{\alpha z} - G_{\alpha z} \zeta_\alpha$.
Further, we have
\begin{align}
\tilde{I}_{\alpha z} 
& = 
I_{\alpha z} 
- F^\mL_{\alpha z} \zeta_\alpha
- \zeta_\alpha F^\mR_{\alpha z}
+ \zeta_\alpha G_{\alpha z} \zeta_\alpha
.
\label{eq:Itilde}
\end{align}
Applying these relations to the two $\Sigma$ estimators yields
\begin{subequations}
\begin{align}
\tilde{\Sigma}^\FG_{\alpha z} 
& =
\tilde{F}^\mL_{\alpha z} (G_{\alpha z})^{-1}
=
F^\mL_{\alpha z} (G_{\alpha z})^{-1} - \zeta_\alpha
,
\\
\tilde{\Sigma}^\IFG_{\alpha z} 
& =
\tilde{\Sigma}^\mH_\alpha
+ \tilde{I}_{\alpha z}
-
\tilde{F}^\mL_{\alpha z} (G_{\alpha z})^{-1} \tilde{F}^\mR_{\alpha z}
\nonumber
\\
& =
\Sigma^\mH_\alpha - \zeta_\alpha
+ I_{\alpha z}
-
F^\mL_{\alpha z} (G_{\alpha z})^{-1} F^\mR_{\alpha z}
.
\label{eq:SigmaIFGtilde}
\end{align}
\end{subequations}
Hence, for an algorithm like NRG, which is bilinear in the arguments of a correlation function $\lala A, B \rara_z$,
a shift does \textit{not} affect the numerical results for $\Sigma^\FG$ and $\Sigma^\IFG$.
For other estimators like $\Sigma^\IF$, involving shifted correlation functions in the \textit{denominator}, 
the equivalence under a shift does not simply follow from linearity but
requires more intricate cancellations that may be violated numerically.
We also note that Eq.~\eqref{eq:Itilde} naturally produces both $F^\mL$ and $F^\mR$.
Hence, for the equivalence of $\Sigma^\IFG$ under shifts according to Eq.~\eqref{eq:SigmaIFGtilde},
it is helpful to use the symmetric form $F^\mL_{\alpha z} G_{\alpha z}^{-1} F^\mR_{\alpha z}$---%
instead of $(F^\mL_{\alpha z})^2 G_{\alpha z}^{-1}$ or $G_{\alpha z}^{-1} (F_{\alpha z}^\mR)^2$---%
if $F^\mL$ and $F^\mR$ (slightly) differ numerically.

Now, even if the shifts leave the numerical results for $\Sigma^\FG$ and $\Sigma^\IFG$ invariant, 
they help us to gain more analytical insight.
Two specific shifts are particularly suited for that.

The first is $\zeta_\alpha = \Sigma^\mH_\alpha$.
With $(\tilde{G}^0_{\alpha z})^{-1} = (G^0_{\alpha z})^{-1} - \Sigma^\mH_\alpha$,
it transforms the bare propagator into the Hartree propagator,
$\tilde{G}^0_{\alpha z} = G^\mH_{\alpha z}$.
This is particularly convenient for particle-hole symmetric systems, 
where $\epsilon_{d,\alpha}$ and $\Sigma^\mH_\alpha$ cancel.
Furthermore, $\tilde{\Sigma}^\mH = 0$ simplifies the $\Sigma$ estimators involving the Hartree self-energy.
One gets, e.g.,
$\tilde{\Sigma}^{\mathrm{IG}}_{\alpha z}
\!=\!
\tilde{I}_{\alpha z} G^\mH_{\alpha z} (G_{\alpha z})^{-1}$,
an estimator used in Ref.~\cite{Enenkel2022}.
Additionally, $\tilde{\Sigma}^\mH = 0$ implies $\tilde{F}_\alpha^{(0)} = 0$
\footnote{In NRG, the value of $\Sigma^\mH_\alpha \!=\! F^{(0)}_\alpha$ may slightly differ among $z$ shifts. Using a $z$-dependent shift $\zeta_\alpha \!=\! \Sigma^\mH_\alpha$ makes each term consistently obey $\tilde{F}^{(0)}_\alpha \!=\! 0$ and slightly improves $\Sigma^\IFG_{\alpha z}$ results.}.
With $\Im \Sigma_{\alpha \nu} = \Im \tilde{\Sigma}_{\alpha \nu}$,
Eqs.~\eqref{eq:SigmaIFG_weight} and \eqref{eq:high-freq_SigmaIFG} then simplify as
\begin{flalign}
\nint \md \nu \, \Im \Sigma_{\alpha \nu}^\IFG
& =
\nint \md \nu \, \Im \tilde{I}_{\alpha \nu}
\ \
\Rightarrow
\ \
\Sigma^{(0)}_\alpha
=
\langle \{ \tilde{q}_\alpha, \tilde{q}_\alpha^\dag \} \rangle
.
\hspace{-0.5cm}
&
\label{eq:SigmaIFG_weight_rephrased}
\end{flalign}

The second interesting shift is 
$\zeta_\alpha = \frac{\Im F_{\alpha \bar{\nu}}}{\Im G_{\alpha \bar{\nu}}}$,
where $\bar{\nu}$ is any given frequency,
as it yields $\Im \tilde{F}_{\alpha \nu}=0$ at $\nu=\bar{\nu}$.
With $\Im \Sigma_{\alpha \nu} = \Im \tilde{\Sigma}_{\alpha \nu}$,
the result of Eq.~\eqref{eq:ImSigmaIFG_derivation}
then simplifies as
\begin{align}
\Im \Sigma^\IFG_{\alpha \bar{\nu}}
=
\Im \tilde{I}_{\alpha \bar{\nu}}
+
\frac{ |G_{\alpha \bar{\nu}}|^2 }
{ \Im G_{\alpha \bar{\nu}} }
( \Im \Sigma^\FG_{\alpha \bar{\nu}} )^2
.
\label{eq:SigmaIFG_lowenergy_rephrased}
\end{align}
Here, the sign of the two summands is determined by
$\Im \tilde{I}_{\alpha \bar{\nu}}$ and $\Im G_{\alpha \bar{\nu}}$, respectively, where
$\tilde{I}_{\alpha \bar{\nu}} \!=\! \lala \tilde{q}_\alpha, \tilde{q}_\alpha^\dag \rara_{\bar{\nu}}$
and $G_{\alpha \bar{\nu}} \!=\! \lala d_\alpha, d_\alpha^\dag \rara_{\bar{\nu}}$.
Since each of them is defined with a mutually conjugate pair of operators,
their Lehmann representations, evaluated with fdm NRG, 
directly yield 
$\Im \tilde{I}_{\alpha \bar{\nu}} \!\leq\! 0$ and $\Im G_{\alpha \bar{\nu}} \!\leq\! 0$,
thus ensuring $\Im \Sigma^\IFG_{\alpha \bar{\nu}} \!\leq\! 0$.
While this analytic argument refers to an arbitrary but fixed frequency $\bar{\nu}$, 
one need not actually perform a shift for each frequency value to 
numerically profit from Eq.~\eqref{eq:SigmaIFG_lowenergy_rephrased}.
Instead, by linearity, NRG results are equivalent for any shift,
and Eq.~\eqref{eq:SigmaIFG_lowenergy_rephrased} ensures $\Im \Sigma^\IFG_{\alpha \nu} \leq 0$
for all frequencies at once.

Interestingly, we find from Eq.~\eqref{eq:SigmaIFG_lowenergy_rephrased}
that the retarded self-energy has a negative imaginary part
without resorting to perturbation theory (which may break down for non-Fermi liquids)
or to properties of the propagator \cite{Luttinger1961}.
Hence, this argument also applies to general quantum impurity models,
for which the retarded nature of 
$G_{\alpha\nu} = 1/(\nu - \epsilon_{d,\alpha} - \Delta_{\alpha \nu} - \Sigma_{\alpha \nu})$
merely requires $\Im (\Delta_{\alpha \nu} + \Sigma_{\alpha\nu}) \leq 0$,
i.e., $\Im \Sigma_{\alpha\nu} \leq -\Im \Delta_{\alpha\nu}$,
instead of $\Im \Sigma_{\alpha\nu} \leq 0$.

\subsection{Denominator in \texorpdfstring{$\Sigma^\IF$}{Sigma-IF}}
\label{sec:denominator}
We mentioned before that
$\Sigma^\IF_{\alpha z}$ is disadvantageous 
since the denominator is problematic for systems with reduced spectral weight
(such as bad metals in DMFT) or even spectral gaps (insulators).
Indeed, let us consider a particle-hole symmetric system,
where $\Re G_{\alpha \nu}$ and $\Re G^\mH_{\alpha \nu}$
are antisymmetric in $\nu$
and thus vanish at $\nu=0$.
From the analog of Eq.~\eqref{eq:1st-order_eom} under the shift $\zeta_\alpha = \Sigma^\mH_\alpha$,
we then have
$G_{\alpha z} = G^\mH_{\alpha z} (1 + \tilde{F}_{\alpha z})$
and
\begin{align}
\Im G_{\alpha, \nu=0} 
= 
\Im G^\mH_{\alpha, \nu=0} \,
(1 + \Re \tilde{F}_{\alpha, \nu=0})
.
\label{eq:ImG_ReF}
\end{align}
Hence, if the spectrum is gapped, $\Im G_{\alpha, \nu=0} = 0$,
Eq.~\eqref{eq:ImG_ReF} shows that
$1 + \Re \tilde{F}_{\alpha, \nu=0} = 0$,
i.e., $\Re \tilde{F}_{\alpha, \nu=0} = -1$.
However, it is numerically challenging to precisely resolve the finite value
to which a Kramers--Kronig transformed object like $\Re \tilde{F}_{\alpha, \nu}$ converges.
For this reason, 
$\Sigma^\IF_{\alpha z}$ is numerically disadvantageous 
for gapped system and, more generally, those with strongly reduced spectral weight
(as demonstrated below).

As a curiosity, we mention that $\Sigma^\IFG_{\alpha \nu}$ can be viewed as a linear interpolation between $\Sigma^\IF_{\alpha \nu}$ and $\Sigma^\FG_{\alpha \nu}$, in the form
\begin{align}
\Sigma^\IFG_{\alpha \nu}
=
\Sigma^\IF_{\alpha \nu} f_\nu
+
\Sigma^\FG_{\alpha \nu} ( 1 - f_\nu )
,
\quad
f_\nu = 1 + F_{\alpha \nu}
.
\end{align}
The weighting function $f_\nu$ is unity for $\nu \to \infty$ and close to unity for $\nu \to 0$ in a Fermi liquid.
Hence, $\Sigma^\IFG_{\alpha \nu}$ and $\Sigma^\IF_{\alpha \nu}$ share many of their beneficial properties 
at high and low energies.
Further, $f_\nu$  is small whenever $1+F_{\alpha \nu}$ is small, i.e., whenever the denominator in $\Sigma^\IF_{\alpha \nu}$ becomes problematic.
In this region, $\Sigma^\IFG_{\alpha \nu}$ is given by $\Sigma^\FG_{\alpha \nu}$ and thus free from any instabilities.

\begin{figure*}[t]
\includegraphics[scale=1]{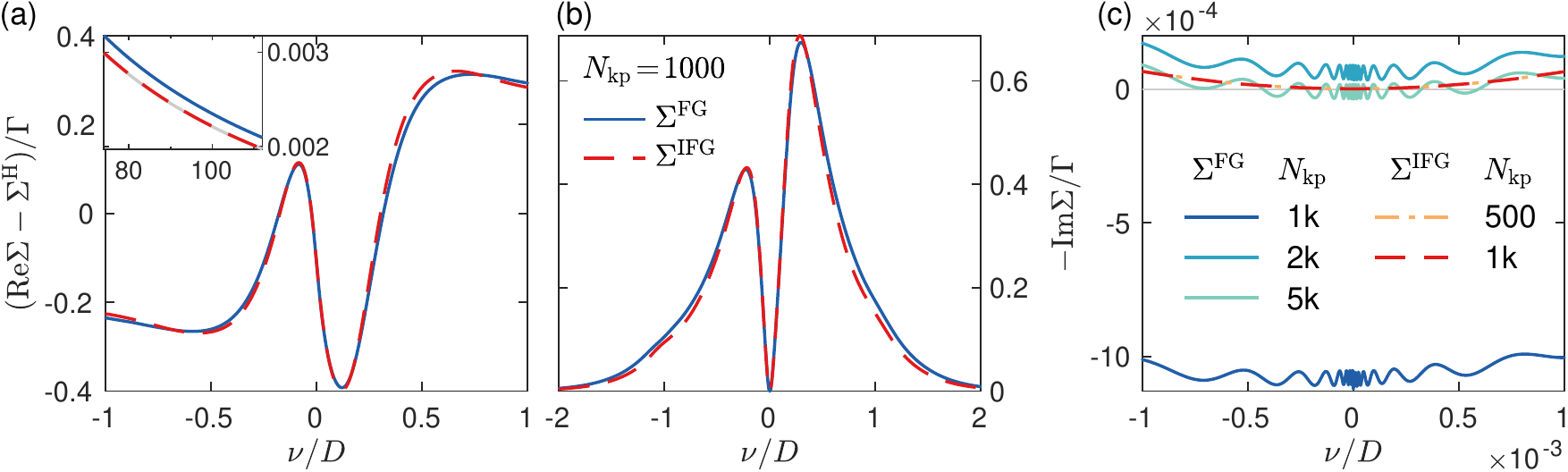}
\caption{%
Self-energies $\Sigma^\FG$ and $\Sigma^\IFG$ for the single-orbital Anderson impurity model.
(a) Real part (minus the Hartree shift), (b) imaginary part. 
The discrepancies in (a) and (b) reflect the fact that $\Sigma^{(0)} \!=\! -\frac{1}{\pi} \nint \md \nu \, \Im \Sigma_\nu$ is exact in $\Sigma^\IFG$ but not in $\Sigma^\FG$.
Indeed, the inset in (a) enlarges the high-frequency decay with the exact asymptote $\Sigma^{(0)}_\ex/\nu$ in gray.
(c) $\Im \Sigma_\nu$ at low energies for an increasing number of kept states $N_\kp$ [SU(2) multiplets; $1\mathsf{k} \!=\! 10^3$, etc.].
Crucially, $-\Im \Sigma^\IFG_\nu$ is nonnegative, free of any wiggles, and, on the given scale, converged for $N_\kp$ as low as $500$.}
\label{fig:SIAM_SE}
\end{figure*}

\section{Numerical results}
\label{sec:numerics}
\subsection{NRG setting}
We employ the fdm NRG \cite{Weichselbaum2007}
in a state-of-the-art implementation based on the 
QSpace tensor library \cite{Weichselbaum2012a,*Weichselbaum2012b,*Weichselbaum2020},
allowing one to exploit Abelian and non-Abelian symmetries.
Indeed, SU(2) spin symmetry is used throughout,
while two calculations additionally have SU(2) charge and SU(3) orbital symmetry, respectively.
The resolution at finite $\nu$ is improved
by averaging the (discrete) spectral data over $n_z$ shifted discretization grids \cite{Zitko2009}
and through an adaptive broadening scheme \cite{Lee2016,Lee2017}.
For the single-orbital results, we set the NRG discretization parameter to $\Lambda \!=\! 2$
and $n_z \!=\! 4$.
For the multiorbital results, we use $\Lambda \!=\! 4$ while fixing $n_z \!=\! 2$ to reduce numerical run times.
The only truncation criterion during the NRG iterative diagonalization is given by the number of kept states $N_\kp$.
As usual, this bound is soft in order to respect emergent degeneracies in the spectrum. 

We will first discuss the Anderson impurity model with a featureless hybridization function.
All other results stem from DMFT solutions of lattice systems
which are mapped onto self-consistently determined impurity models.
For simplicity, we consider the Bethe lattice with a semicircular lattice density of states
and converge the DMFT self-consistency iteration using the conventional self-energy scheme $\Sigma^\FG$.
Although the self-consistency condition on the Bethe lattice can be phrased in terms of the spectral function 
$\Ac_{\alpha \nu} = - \tfrac{1}{\pi} \Im G_{\alpha \nu}$,
it is standard practice to use the self-energy for an improved spectral function compared to the direct NRG output.
Then, from the converged DMFT solution, we perform one more calculation to compare results with different self-energy estimators.

Generally, we use the half bandwidth $D=1$ of the (bare) hybridization function as our energy unit.
For the plain Anderson impurity model, the size of $\Sigma$ is best compared to the hybridization strength $\Gamma$.
In lattice systems, the self-energy adds to the dispersion relation in the inverse propagator; we thus plot $\Sigma/D$.
For all our numerical results, we set $T/D = 10^{-8}$, which can be considered as zero temperature.

\subsection{Single-orbital Anderson impurity model}
\label{sec:SIAM}
We begin our presentation of numerical results with the 
single-orbital Anderson impurity model
[cf.\ Eq.~\eqref{eq:Anderson}],
with a box-shaped hybridization function
$- \Im \Delta_\nu = \Gamma \theta(D - |\nu|)$
of half bandwidth $D=1$ and strength $\Gamma$.
We here choose $\Gamma=0.1$, 
an interaction value of $U=0.3$, 
and the on-site energy $\epsilon_d = -0.1$.
With $\epsilon_d > -U/2$, the system is less than half filled,
having $n_d = \sum_\sigma \langle n_\sigma \rangle \approx 0.87$.
Appendix~\ref{sec:appendix2} provides additional results,
obtained for the same parameter set as chosen in Ref.~\cite{Bulla1998}:
$\Gamma=0.015$, $U=0.2$, and $\epsilon_d = -0.1$ (half filling).
For the present model,
$q_\sigma = U d_\sigma n_{\bar{\sigma}}$.
Hence, the Hartree self-energy gives the well-known 
$\Sigma^\mH_\sigma = \langle \{ q_\sigma, d^\dag_\sigma \} \rangle = U n_d/2$.
Using Eq.~\eqref{eq:SigmaIFG_weight_rephrased}, the (exact) self-energy moment is easily evaluated as 
$\Sigma^{(0)}_\ex \!=\! \langle \{ \tilde{q}_\sigma, \tilde{q}^\dag_\sigma \} \rangle \!=\! U^2 \tfrac{n_d}{2}(1 \!-\! \tfrac{n_d}{2})$.
Our temperature $T \!=\! 10^{-8}$ is far below the Kondo temperature
\footnote{Here, we use the well-known formula \cite{Haldane1978,Merker2012} for the Kondo temperature
$T_\mK = (U\Gamma/2)^{1/2} \exp[ \pi \epsilon_d (\epsilon_d+U)/(2U\Gamma) ]$.}
\nocite{Haldane1978,Merker2012}
of $T_\mK \approx 0.043$. 

To set the stage, Figs.~\ref{fig:SIAM_SE}(a) and \ref{fig:SIAM_SE}(b) show the real and imaginary part of $\Sigma$ on a wide energy window.
Generally, $\Sigma^\FG_\nu$ and $\Sigma^\IFG_\nu$ yield consistent results, 
while slight deviations are observed at higher energies. 
This is expected since
$\Sigma^\IFG_\nu$ gives the exact high-frequency asymptote for the real part [see inset of Fig.~\ref{fig:SIAM_SE}(a)] and the exact total weight for the imaginary part,
whereas both properties are slightly violated in $\Sigma^\FG_\nu$.
Before inspecting this further, we enlarge the low-energy behavior of $\Im \Sigma_\nu$ in Fig.~\ref{fig:SIAM_SE}(c)
and compare results for different numbers of kept states $N_\kp$ [here SU(2) multiplets].
For $N_\kp = 1000$,
$-\Im \Sigma^\FG_\nu$ overshoots to negative values on the scale of $10^{-3}\Gamma$.
Increasing $N_\kp$, $-\Im \Sigma^\FG_\nu$ approaches the $\nu$ axis.
Importantly, however, the notable wiggles of $\Im \Sigma^\FG_\nu$ on the scale of $10^{-4}\Gamma$ remain, even for $N_\kp$ as high as $5000$.
In striking contrast, $-\Im \Sigma^\IFG_\nu$ is nonnegative, free of any wiggles, and, on the given scale, already converged for $N_\kp$ as low as $500$.

\begin{figure}[t]
\includegraphics[scale=1]{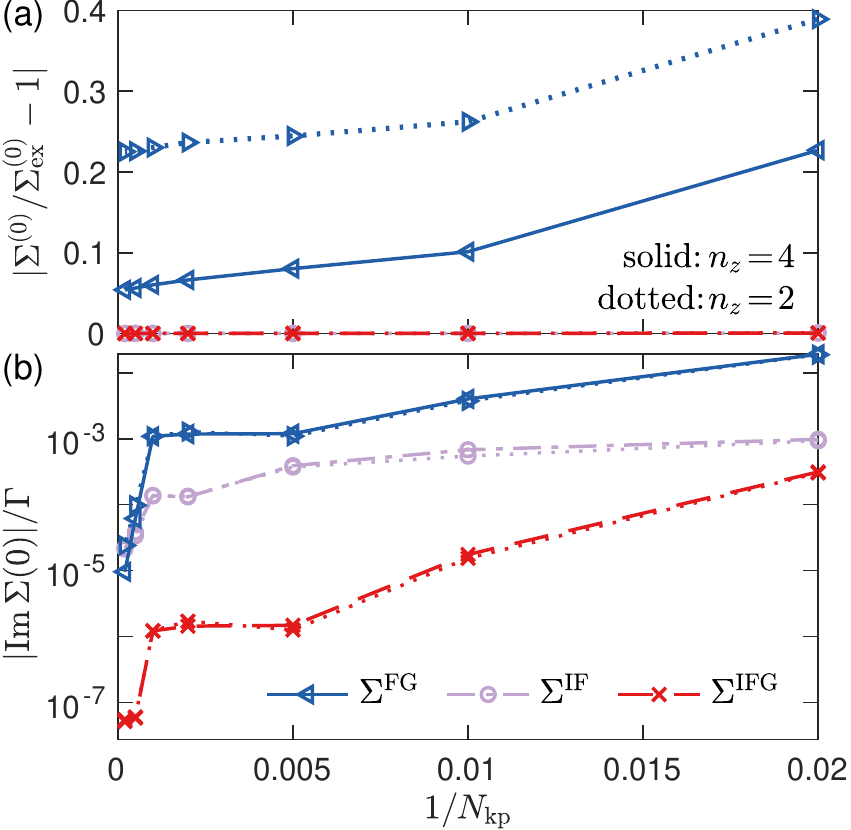}
\caption{%
The self-energy moment $\Sigma^{(0)} \!=\! - \frac{1}{\pi} \nint \md \nu \, \Im \Sigma_\nu$ and $\Im \Sigma_{\nu=0}$ as a function of $1/N_\kp$.
(a) The $\Sigma^\FG$ results improve with increasing $N_\kp$ and $n_z$, the number of $z$ shifts;
$\Sigma^\IFG$ (and $\Sigma^\IF$) always yield the exact value.
(b) All results get closer to the exact value zero with increasing $N_\kp$ while being almost independent of $n_z$.
Throughout, $\Sigma^\IFG$ improves upon $\Sigma^\FG$ by two to three orders of magnitude.}
\label{fig:SIAM_Nk}
\end{figure}

In Fig.~\ref{fig:SIAM_Nk}, we take a closer look at both the high-energy property $\Sigma^{(0)} = - \tfrac{1}{\pi} \nint \md \nu \, \Im \Sigma_\nu$ and the low-energy property $\Im \Sigma_{\nu=0}$.
The former should give 
$\Sigma^{(0)}_\ex = U^2 \tfrac{n_d}{2}(1 - \tfrac{n_d}{2})$,
the latter zero at $T=0$ [or more generally $\mathit{O}(T^2/T_\mK$)].
We plot both quantities as a function of $1/N_\kp$, i.e., with accuracy increasing toward the left.
Starting with Fig.~\ref{fig:SIAM_Nk}(a) and $\Sigma^\FG$, we see that the deviation of $\Sigma^{(0)}$ from $\Sigma^{(0)}_\ex$ decreases with increasing the number of kept states, $N_\kp$, and the number of $z$ shifts $n_z$.
However, the relative deviation stays above $5\%$ even for the high-accuracy setting $N_\kp = 5000$ and $n_z = 4$.
By contrast, for $\Im \Sigma^\IFG_\nu$ (and also $\Im \Sigma^\IF_\nu$),
the exact total weight is guaranteed, no matter the chosen parameters.

As a low-energy property,
$\Im \Sigma_{\nu=0}$, shown in Fig.~\ref{fig:SIAM_Nk}(b),
is basically independent of discretization details and thus almost the same for $n_z = 2$ and $n_z = 4$. 
One readily observes that $\Im \Sigma^\IFG_{\nu=0}$ is more accurate than $\Im \Sigma^\FG_{\nu=0}$ 
(and also $\Im \Sigma^\IF_{\nu=0}$)
by several orders of magnitude.
In more detail,
for all self-energy estimators, the values improve (quasi) monotonically with $N_\kp$ and approach the exact value, zero.
For the lower values of $N_\kp$, the numerical error in $\Im \Sigma^\IFG_{\nu=0}$
is dominated by $(\Im \Sigma^\FG_{\nu=0})^2$ in Eq.~\eqref{eq:ImSigmaIFG_derivation},
leading to the quadratic relation between both curves in Fig.~\ref{fig:SIAM_Nk}(b).
For higher $N_\kp$ and extremely low values of $|\Im \Sigma^\IFG_{\nu=0}|$,
uncertainties in the first two terms of Eq.~\eqref{eq:ImSigmaIFG_derivation} become noticeable.
Indeed, at $N_\kp = 1000$, e.g., $\Sigma^\FG$, $\Sigma^\IF$, and $\Sigma^\IFG$
roughly yield $10^{-3}$, $10^{-4}$, and $10^{-6}$, respectively.
Increasing $N_\kp$ to $5000$, $-\Im \Sigma^\FG_{\nu=0}/\Gamma$ reaches down to $10^{-5}$ and
$-\Im \Sigma^\IFG_{\nu=0}/\Gamma$ even down to $10^{-7}$.

\subsection{Single-orbital Hubbard model}
\label{sec:Hubbard}
Next, we consider the DMFT+NRG solution
of the single-orbital Hubbard model 
(Bethe lattice with half bandwidth $D=1$).
We set the interaction value to $U=2.6$ in the metal--insulator coexistence region.
Metallic and insulating solutions are obtained by approaching $U$ from below and above, respectively,
in the DMFT self-consistency iteration.
For this particle-hole symmetric setup, we exploit SU(2) charge and SU(2) spin symmetry,
keeping $N_\kp = 5000$ multiplets.
Working with the shift $\zeta = \Sigma^\mH$ (cf.\ Sec.~\ref{sec:shift}),
we particularly have $\tilde{\Sigma}^\IF_z = \tilde{I}_z / (1+\tilde{F}_z)$.

\begin{figure}[t]
\includegraphics[scale=1]{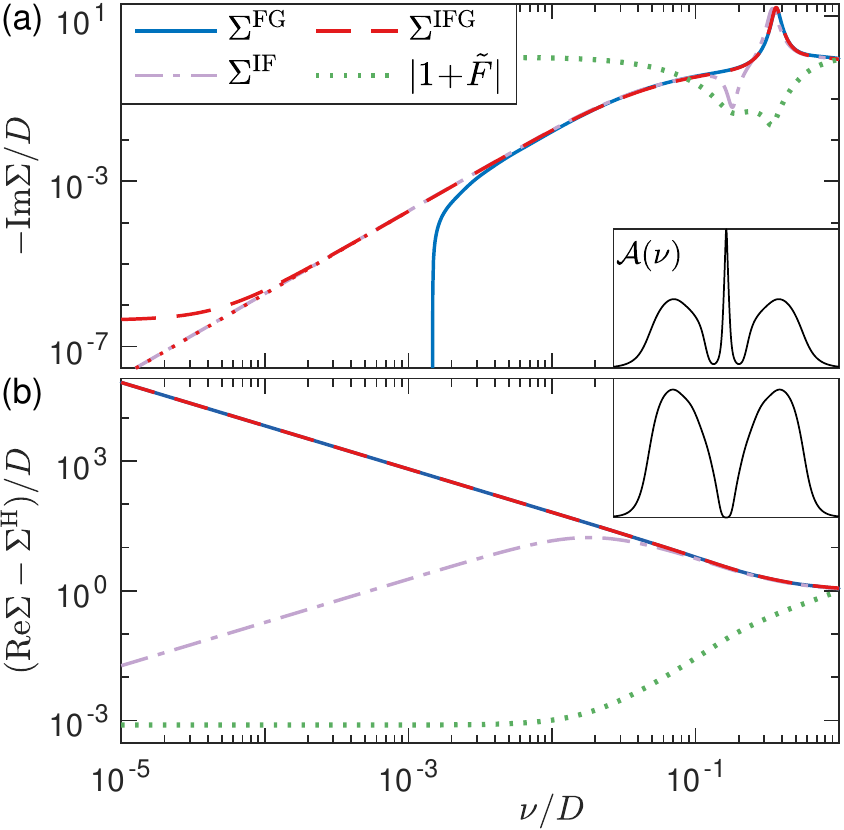}
\caption{%
Self-energies for the single-orbital Hubbard model
in (a) the metallic and (b) the insulating phase 
(see spectral functions in the insets)
at $U/D \!=\! 2.6$.
(a) While $-\Im \Sigma^\FG_\nu$ overshoots to negative values and $-\Im \Sigma^\IF_\nu$ has artifacts when $|1 \!+\! \tilde{F}_\nu|$ becomes small, $-\Im \Sigma^\IFG_\nu$ gives reliable results throughout.
The red dotted line shows the low-energy behavior via Eq.~\eqref{eq:SigmaIFG_lowenergy} and follows the $\nu^2$ decay smoothly down to $-\Im \Sigma^\IFG_\nu/D \!\sim\! 10^{-13}$.
(b) $\Re \Sigma^\FG_\nu$ and $\Re \Sigma^\IFG_\nu$ follow the $1/\nu$ divergence of the Mott insulator,
but $\Re \Sigma^\IF_\nu$ deviates from that as soon as $|1 \!+\! \tilde{F}_\nu|$ levels off, due to inaccuracies in $\Re \tilde{F}_\nu$.}
\label{fig:HM_SE}
\end{figure}

Figures~\ref{fig:HM_SE}(a) and \ref{fig:HM_SE}(b) show $\Im \Sigma_\nu$ for the metallic solution
and $\Re \Sigma_\nu$ for the insulating solution, respectively.
Overall, $\Sigma^\FG$ and $\Sigma^\IFG$ give consistent results for both phases.
For the metal, however, $-\Im \Sigma^\FG_\nu$ overshoots to unphysical negative values at the point $\nu \approx 2 \cdot 10^{-3}$ and $-\Im \Sigma^\FG_\nu \approx 10^{-3}$, while $-\Im \Sigma^\IFG_\nu$ decreases smoothly down to $-\Im \Sigma_\nu \approx 10^{-6}$.
Again, the two values are related quadratically by Eq.~\eqref{eq:ImSigmaIFG_derivation}.
The red dotted line in the plot gives the low-energy behavior of $\Im \Sigma^\IFG_\nu$
according to Eq.~\eqref{eq:SigmaIFG_lowenergy}, i.e.,
by discarding the erroneous $(\Im \Sigma^\FG_\nu)^2$ in Eq.~\eqref{eq:ImSigmaIFG_derivation}.
For this highly symmetric (and highly accurate) calculation, 
the result goes smoothly down in energy, all the way to $-\Im \Sigma^\IFG_\nu/D \!\sim\! 10^{-13}$.

Turning to $\Sigma^\IF$ in Fig.~\ref{fig:HM_SE}(a), 
the curve also shows a clean $\nu^2$ decay, but it has
a peculiar dip at larger frequencies, $\nu \approx 0.2$.
The reason is that the spectral function in the metallic phase (see inset)
exhibits strongly reduced weight at precisely these frequencies, separating the quasiparticle peak from the Hubbard bands.
This leads to low values in $|1 + \tilde{F}_\nu|$ (as explained in Sec.~\ref{sec:denominator}) and artifacts in $\Sigma^\IF_\nu$.
In the insulating phase, Fig.~\ref{fig:HM_SE}(b), 
$\Re \Sigma^\FG$ and $\Re \Sigma^\IFG$ nicely follow the $1/\nu$ divergence of the Mott insulator.
There, $|1 + \tilde{F}_\nu|$ should decrease to zero for $\nu \to 0$. 
However, due to inaccuracies in the real part obtained by Kramers--Kronig transform, $|1 + \tilde{F}_\nu|$ levels off for $\nu < 0.01$, so that $\Re \Sigma^\IF_\nu$ deviates from $1/\nu$.
In total, for this strongly correlated setup, $\Sigma^\IF$ does \textit{not} give reliable results
since the denominator $1+\tilde{F}_\nu$ leads to numerical instabilities.

\subsection{Multiorbital Hubbard models}
\begin{figure}[t]
\includegraphics[scale=1]{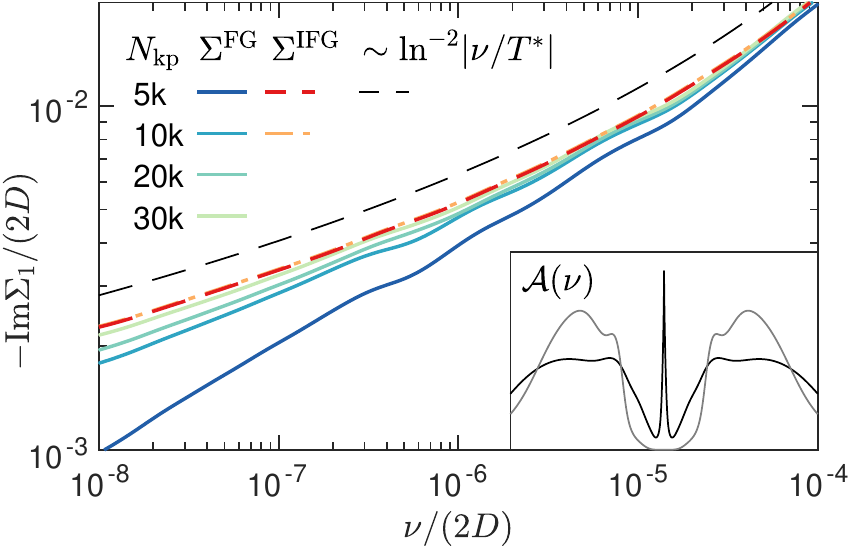}
\caption{%
Self-energies $\Sigma_1$ of the metallic orbital in a two-orbital realization of the OMSP
(see spectral functions in the inset).
As a singular Fermi liquid,
$-\Im \Sigma_{1,\nu}$ is expected to show a logarithmic singularity.
With $\Sigma^\FG$, this behavior is only seen for the largest $N_\kp$.
By contrast, $\Im \Sigma^\IFG_{1,\nu}$ perfectly agrees with the analytic curve
and is, on the given scale, already converged for the lowest $N_\kp = 5000$.
The extra factor $2$ on the axes is used to have convenient tick marks.}
\label{fig:OSMP_SE}
\end{figure}

We now turn to DMFT+NRG results for multiorbital Hubbard models 
on the Bethe lattice with the interaction given by Eq.~\eqref{eq:Dworin}.
We first consider two half-filled orbitals and different bandwidths of $D \equiv D_1=1$ and $D_2=0.5$.
Setting $U=1.8$ and $J=0.3$ yields a simple realization of an orbital-selective Mott phase (OSMP).
Indeed, in the absence of interorbital hopping \cite{Kugler2021}, 
the 1-orbital is metallic while the 2-orbital is a Mott insulator with a gap in the spectrum
(see inset of Fig.~\ref{fig:OSMP_SE}).
Note that, for the present calculation, the Wilson chains of the two orbitals are interleaved 
\cite{Mitchell2014,Stadler2016} for extra efficiency.

The main panel of Fig.~\ref{fig:OSMP_SE} shows the self-energy of the metallic 1-orbital, $\Sigma_1$, at low energies
and for various $N_\kp$.
It is known that,
due to the unscreened magnetic moments of the 2-orbital,
the 1-orbital is a singular Fermi liquid with logarithmic singularities in $\Sigma_1$
\cite{Greger2013,Kugler2019}.
The black dashed line shows the expected $a\ln^{-2}|\nu/T^*|$ behavior;
where $a$ and $T^*$ are fitting parameters.
We multiplied $a$ by an extra factor $1.2$ to separate the curves for better readability.
The numerical data obtained with $\Sigma^\FG$ shifts notably by increasing $N_\kp$
from 5000 to 30000 SU(2) multiplets.
For the lowest accuracy, $\Im\Sigma^\FG_{1,\nu}$ does not reproduce the analytically expected behavior
and still contains notable wiggles.
Increasing $N_\kp$, $\Im\Sigma^\FG_{1,\nu}$ approaches the expected behavior but is not fully converged
at $N_\kp = 30000$.
By contrast, $\Sigma^\IFG$ always yields a stable curve, in excellent agreement with the analytic form, and 
is, on the given scale, already converged for the lowest number of kept states, $N_\kp = 5000$.

\begin{figure}[t]
\includegraphics[scale=1]{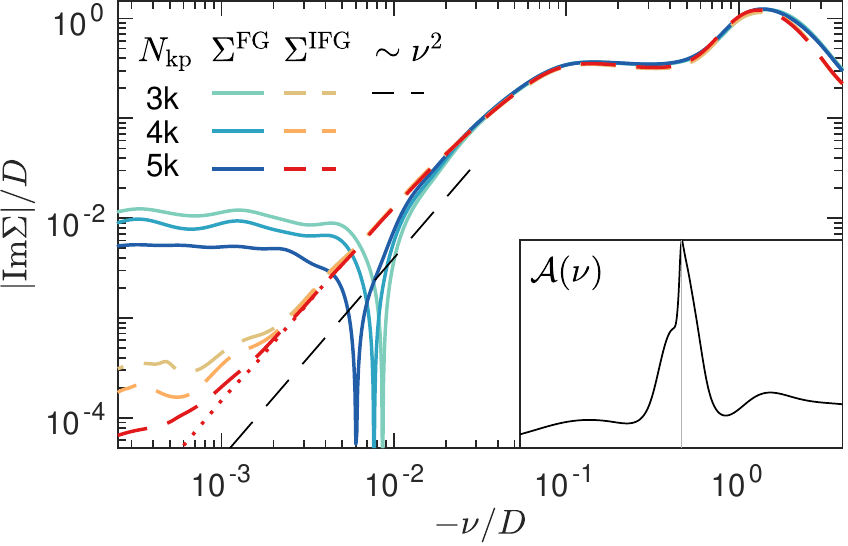}
\caption{%
Self-energies in a degenerate three-orbital Hund-metal system.
We focus on negative frequencies where the spectral function at filling $2$ has a pronounced shoulder (see inset).
$- \Im \Sigma^\FG_\nu$ overshoots to unphysical negative values for $-\nu/D \!\lesssim\! 10^{-2}$
for all numerically feasible $N_\kp$ [SU(2)$\times$SU(3) multiplets].
By contrast, $- \Im \Sigma^\IFG_\nu$ never overshoots and follows the $\nu^2$ decay down to 
$- \Im \Sigma^\IFG_\nu/D \!=\! 10^{-3}$ and even $10^{-4}$ for $N_\kp \!=\! 3000$ and $5000$, respectively.
The red dotted line shows the low-energy behavior according to Eq.~\eqref{eq:SigmaIFG_lowenergy} and follows the $\nu^2$ decay smoothly down to $-\Im \Sigma^\IFG_\nu/D \sim 10^{-6}$.}
\label{fig:Hund_SE}
\end{figure}

We close our presentation of numerical results with a three-orbital setup of the Hund-metal category \cite{Haule2009,Georges2013}.
Having three degenerate orbitals, we exploit the additional SU(3) symmetry permitted by Eq.~\eqref{eq:Dworin}.
There, we set $U=3$, $J=0.5$ in units of the half bandwidth $D=1$.
At a filling of two, the spectral function (see inset of Fig.~\ref{fig:Hund_SE}) is highly asymmetric.
The left part of $\Ac_\nu$ exhibits an intriguing orbital-resonance shoulder \cite{Stadler2016,Stadler2019,Kugler2019},
and we thus focus on $\nu<0$ for the analysis of $\Sigma$.

Figure~\ref{fig:Hund_SE} shows $|\Im \Sigma_\nu|$ for both estimators $\Sigma^\FG$ and $\Sigma^\IFG$ 
and for $N_\kp=3000$, $4000$, and $5000$ SU(2)$\times$SU(3) multiplets.
In this challenging, three-orbital setup, $-\Im \Sigma^\FG_\nu$ overshoots to unphysical negative values
already at the point where $-\nu$ and $-\Im \Sigma$ are around $10^{-2}$.
For $\Sigma^\FG$ calculations with higher $N_\kp$, 
this point is shifted only marginally to lower $\nu$.
Wiggles in $\Im \Sigma^\FG_\nu$ are on the scale of $10^{-3}$ for the lower $N_\kp$
and weaker but still present for the largest $N_\kp$.
Again, $\Sigma^\IFG$ eradicates the overshooting problem.
Even for the lowest $N_\kp$,
$-\Im \Sigma^\IFG_\nu$ follows a clean $\nu^2$ decay down to values of $10^{-3}$,
before wiggles appear on the scale of $10^{-4}$.
For the highest $N_\kp = 5000$, $-\Im \Sigma^\IFG_\nu$ follows the $\nu^2$ decay down to values of $10^{-4}$
and hardly any wiggles are to be found.
Again, the values of $|\Im \Sigma^\FG_{\nu=0}|$ and $|\Im \Sigma^\IFG_{\nu=0}|$
relate quadratically [Eq.~\eqref{eq:ImSigmaIFG_derivation}].
The red dotted line shows $\Im \Sigma^\IFG_\nu$ for $|\nu| \to 0$ according to Eq.~\eqref{eq:SigmaIFG_lowenergy}
and reaches down to values of $-\Im \Sigma^\IFG_\nu/D \sim 10^{-6}$.

It is clear from the plot 
that converging the standard estimator
$\Sigma^\FG$ with $N_\kp$ 
toward a clean $\nu^2$ decay down to, say, $10^{-3}$ is very slow
and practically unfeasible.
By contrast, $\Sigma^\IFG$ gives very accurate results already for much lower $N_\kp$,
and its low-energy behavior can be extracted very cleanly from Eq.~\eqref{eq:SigmaIFG_lowenergy}.
We hope that, in this way, our new self-energy estimator will expand the class of systems
where NRG can be used as a highly-accurate, real-frequency DMFT impurity solver.

\section{Conclusion}
\label{sec:conclusion}
We presented a new self-energy estimator and showed that it yields greatly improved results in NRG calculations.
The standard estimator $\Sigma^\FG=F/G$
[see Eqs.~\eqref{eq:def_G_I_F} and \eqref{eq:SigmaFG_1}]
follows from an eom for $G$.
While it yields much better results than employing the Dyson equation,
$\Sigma^\FG$ still does not have all the qualities one is used to from fdm NRG correlators, as
$\Im \Sigma^\FG$ is not properly normalized and can overshoot to positive values.
Moreover, it often displays wiggles for very low energies.

By combining the eom for $G$ with an analogous eom for $F$, one can derive several $\Sigma$ estimators [see Eqs.~\eqref{eq:SigmaIG}--\eqref{eq:SigmaIF}].
We identified $\Sigma^\IFG = \Sigma^\mH + I - F^2/G$ as particularly well-suited for NRG.
Indeed, we showed analytically that the normalization of $\Im \Sigma^\IFG$ [Eq.~\eqref{eq:SigmaIFG_weight}] 
and its low-energy behavior [see Eq.~\eqref{eq:SigmaIFG_lowenergy}]
are determined directly by $\Im I$, $\Im F$, and $\Im G$.
Accordingly, there are no real parts (obtained by Kramers--Kronig transform) involved,
and $\Im \Sigma^\IFG$ is as reliable as the imaginary part of any fdm NRG correlator:
it is normalized, does not overshoot, and is extremely accurate at low energies.

We examined numerical results for the Anderson impurity model with a featureless hybridization and for the DMFT solutions of one-, two-, and three-orbital Hubbard models.
In all cases, the above properties were confirmed and $\Sigma^\IFG$ yielded much better results than $\Sigma^\FG$. Furthermore, we found that $\Sigma^\IFG$ converged much faster with increasing the numerical effort (increasing the number of kept states $N_\kp$) than $\Sigma^\FG$.
This is very important when applying NRG to multiorbital systems where
the maximal $N_\kp$ is limited by numerical resources and finding accurate results for $\Sigma^\FG$ was the major challenge.

The estimator $\Sigma^\FG=F/G$ is also frequently used for other impurity solvers,
such as exact diagonalization (ED) \cite{Enenkel2022},
the density-matrix renormalization group (DMRG) \cite{Zhu2017,Karp2020}
and quantum Monte Carlo (QMC) \cite{Hafermann2012,Hafermann2013,Hafermann2014,Gunacker2016}.
Although our analysis is targeted at NRG applications, 
we expect that $\Sigma^\IFG$ yields improved results for some of these methods, too.

\section*{Acknowledgments}
The author would like to thank N.~Enenkel for bringing the use of higher-order eoms in ED calculations to his attention,
and J.~von Delft, A.~Gleis, Seung-Sup B.~Lee, and A.~Weichselbaum for helpful comments on the manuscript.
The NRG results were obtained using the QSpace tensor library developed by A.~Weichselbaum 
\cite{Weichselbaum2012a,*Weichselbaum2012b,*Weichselbaum2020}
and the NRG toolbox by Seung-Sup B.~Lee \cite{Lee2016,Lee2017}.
Support by the NSF Grant No.~DMR-1733071 and the Alexander von Humboldt Foundation through the Feodor Lynen Fellowship is gratefully acknowledged. 

\appendix
\section{Matrix-valued correlation functions}
\label{sec:appendix1}
In the main text, we focused on $\alpha$-diagonal fermionic correlation functions,
as they follow from Eqs.~\eqref{eq:H0}--\eqref{eq:Dworin}.
Matrix-valued correlation functions are obtained if the quadratic Hamiltonian is generalized to
\begin{align}
H_0
& =
\textstyle
\sum_{\alpha\alpha'}
d^\dag_{\alpha'} \epsilon_{d,\alpha'\alpha} d_\alpha
+
\sum_{k,\alpha\alpha'}
c^\dag_{k\alpha'} \epsilon_{k,\alpha'\alpha} c_{k\alpha}
\nonumber
\\
& \ 
\textstyle
+
\sum_{k,\alpha\alpha'} ( d_{\alpha'}^\dag V_{k,\alpha'\alpha} c_{k\alpha} + \mathrm{H.c.} )
\label{eq:H0_matrix}
\end{align}
and has nonzero off-diagonal ($\alpha \!\neq\! \alpha'$) elements.
Indeed, this expression contains several matrices, which we denote by $\epsilon_d$, $\epsilon_k$, and $V_k$
without subscripts $\alpha$.
Most results of the main text are purposefully phrased in such a way that they directly generalize to matrix form
upon removing $\alpha$ indices.
An exception is the hybridization function in Eq.~\eqref{eq:G0_Delta}, which is rephrased as
\begin{align}
\textstyle
\Delta_z 
= 
\sum_k V_k (z-\epsilon_k)^{-1} V^\dag_k
.
\end{align}
The matrix-valued correlation functions are denoted by $G_z$, $I_z$, etc.,
without $\alpha$ indices.
Their matrix elements are defined by
\begin{flalign}
\quad
G_{\alpha\alpha', z} 
& = 
\lala d_\alpha, d_{\alpha'}^\dag \rara_z
, \quad
&
I_{\alpha\alpha', z} 
& = 
\lala q_\alpha, q_{\alpha'}^\dag \rara_z
,
&
\\
\quad
F^\mL_{\alpha\alpha',z} 
& =
\lala q_\alpha, d_{\alpha'}^\dag \rara_z
,
\quad
&
F^\mR_{\alpha\alpha', z} 
& =
\lala d_\alpha, q_{\alpha'}^\dag \rara_z
.
&
\end{flalign}

In this generalized setting, too,
one computes with NRG the Lehmann representation of 
a spectral function like
\begin{align}
\Ac^{\mathrm{G}}_\nu
& =
(G_\nu - G_\nu^\dag)/(-2\pi\mi)
.
\end{align}
Its diagonal elements fulfill the standard relation
$\Ac^{\mathrm{G}}_{\alpha\alpha,\nu} \!=\! -\frac{1}{\pi} \Im G_{\alpha \alpha,\nu}$,
while the off-diagonal elements are generally complex.
From the Lehmann representation, one also finds
$\Ac^{\mathrm{F}^\mL}_\nu \!=\! (\Ac^{\mathrm{F}^\mR}_\nu)^\dag$
as the generalization of 
$F^\mL_{\alpha\alpha,z} \!=\! F^\mR_{\alpha\alpha,z}$ used in the main text.
The retarded correlator subsequently follows as
\begin{align}
G_\nu 
= 
\, \mathcal{P} \! \nint \md \nu'
\frac{1}{\nu - \nu'}
\Ac^{\mathrm{G}}_{\nu'}
\, - \,
\mi\pi \Ac^{\mathrm{G}}_\nu
.
\end{align}

\begin{figure}[t]
\includegraphics[scale=1]{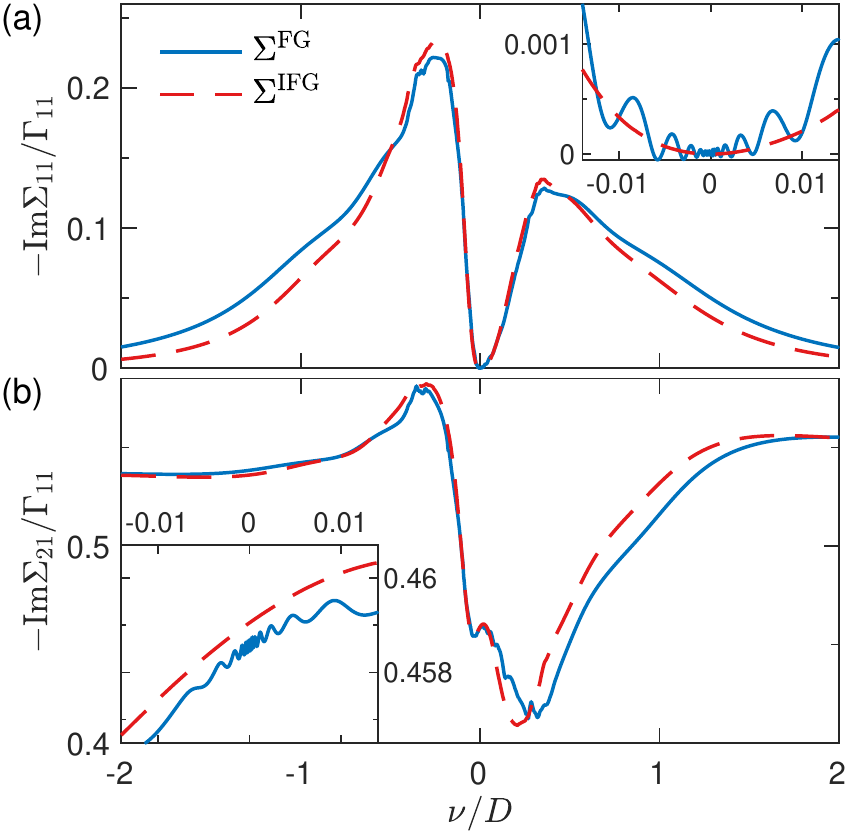}
\caption{%
Imaginary part of two self-energy components for the Anderson impurity model
with off-diagonal on-site energy and hybridization.
For both a diagonal (a) and off-diagonal (b) element of $\Sigma$,
the results from $\Sigma^\IFG$ are smoother than those of $\Sigma^\FG$
and free from wiggles at low energies.
The total weight $\int \md \nu \Ac^\Sigma_\nu$ from $\Sigma^\IFG$ is in perfect agreement with the exact value
$\langle \{ \tilde{q}_\alpha, \tilde{q}^\dag_{\alpha'} \} \rangle$,
cf.\ Eq.~\eqref{eq:SigmaIFG_weight_rephrased},
whereas that obtained with $\Sigma^\FG$ deviates by roughly 15\% in each component.}
\label{fig:SIAM_matrix}
\end{figure}

\begin{figure*}[t]
\includegraphics[scale=1]{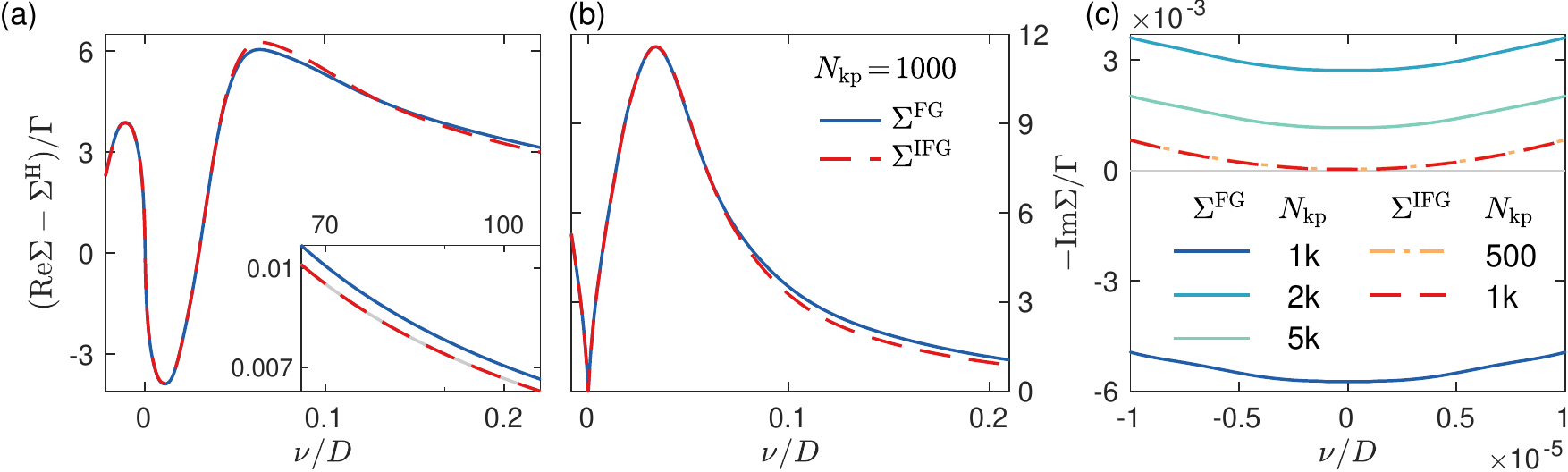}
\caption{%
Self-energies $\Sigma^\FG$ and $\Sigma^\IFG$ for the single-orbital Anderson impurity model,
as in Fig.~\ref{fig:SIAM_SE}, but at a stronger interaction, $\Gamma/U = 0.075$,
and half filling.
(a) Real part (minus the Hartree shift) and (b) imaginary part
on a wide frequency window.
Discrepancies reflect the fact that $\Sigma^{(0)}$ is exact in $\Sigma^\IFG$ but not in $\Sigma^\FG$.
We restrict the frequency range to $\nu > 0$ in light of particle-hole symmetry.
The inset in (a) enlarges the high-frequency decay with $\Sigma^{(0)}_\ex/\nu$ in gray.
(c) $\Im \Sigma_\nu$ at low energies for an increasing number of kept states $N_\kp$ [SU(2)$\times$SU(2) multiplets].
One observes that $-\Im\Sigma^\FG_\nu$ overshoots to negative values for $N_\kp=1000$ 
and is not yet converged for the highest $N_\kp=5000$.
By contrast, $-\Im\Sigma^\IFG_\nu$ shows a clean, nonnegative parabola and is, on the given scale,
converged for $N_\kp$ as low as $500$.}
\label{fig:SIAM_SE_largeU}
\end{figure*}

The equations of motion \eqref{eq:1st-order_eom} and \eqref{eq:2nd-order_eom} involving $F^\mL$,
together with their counterparts involving $F^\mR$,
were already given in a way that directly generalizes to matrix form.
The same applies to the $\Sigma$ formulas \eqref{eq:SigmaFG}--\eqref{eq:SigmaIF}
if $F$ without superscript is understood as $F^\mL$.
Here, we gather the various matrix-valued $\Sigma$ estimators in both their ``left'' and ``right'' forms:
\begin{subequations}
\begin{align}
\label{eq:SigmaFG_matrix}
\Sigma^\FG_z
& =
F_z^\mL G_z^{-1}
\\
& =
G_z^{-1} F_z^\mR
,
\\
\Sigma^{\mathrm{IG}}_z
& =
(\Sigma^\mH + I_z ) G^0_z G_z^{-1}
\\
& =
G_z^{-1} G^0_z ( \Sigma^\mH + I_z )
,
\\
\Sigma^{\mathrm{I}}_z
& =
[1 + ( \Sigma^\mH + I_z ) G^0_z ]^{-1} (\Sigma^\mH + I_z)
\\
& =
(\Sigma^\mH + I_z) [1 + G^0_z ( \Sigma^\mH + I_z )]^{-1}
,
\\
\Sigma^\IF_z
& =
(1 + F^\mL_z)^{-1} (\Sigma^\mH + I_z)
\\
& =
(\Sigma^\mH + I_z) (1 + F^\mR_z)^{-1}
,
\\
\Sigma^\IFG_z
& =
\Sigma^\mH + I_z -
F^\mL_z G_z^{-1} F^\mR_z
.
\label{eq:SigmaIFG_matrix}
\end{align}
\end{subequations}
Note that the last term can also be written as 
$(F^\mL_z)^2 G_z^{-1}$,
$G_z^{-1} (F^\mR_z)^2$.
However, this spoils the invariance under shifts (see Sec.~\ref{sec:shift})
and is therefore numerically disadvantageous.

Finally, we present an exemplary set of numerical results for matrix-valued self-energies.
We consider an Anderson impurity model, similar to the one from Sec.~\ref{sec:SIAM},
with a boxed-shaped hybridization function 
$\Ac^\Delta_\nu = \Gamma \theta(D - |\nu|) / \pi$,
and interaction strength $U=0.3$ and temperature $T=10^{-8}$ in units of the half bandwidth $D=1$.
However, differently from Sec.~\ref{sec:SIAM}, we promote the on-site energy $\epsilon_d$ and the 
hybridization strength $\Gamma$ to non-diagonal matrices:
\begin{align}
\epsilon_d
=
\begin{pmatrix}
-0.2 & 0.05\mi \\
-0.05\mi & -0.1
\end{pmatrix}
, \qquad
\Gamma
=
\begin{pmatrix}
1 & -\mi \\
\mi & \phantom{+}1
\end{pmatrix}
.
\end{align}

This model exhibits only a U(1) charge symmetry,
and we choose the NRG parameters as $\Lambda=2$, $n_z=2$, $N_\kp = 8000$.
The self-energies $\Sigma^\FG$ and $\Sigma^\IFG$ are obtained from their matrix expressions \eqref{eq:SigmaFG_matrix} and \eqref{eq:SigmaIFG_matrix}.
Figure~\ref{fig:SIAM_matrix} shows the imaginary part of two exemplary components, $\Sigma_{11}$ and $\Sigma_{21}$.
Overall, both estimators yield consistent results.
However, already on the wide frequency window, one observes that $\Sigma^\IFG$ results are much smoother than those of $\Sigma^\FG$.
The notable differences at large energies owe to the fact that $\Sigma^\IFG$ produces the exact total weight
$\int \md \nu \Ac^\Sigma_\nu$ as known from $\langle \{ \tilde{q}_\alpha, \tilde{q}^\dag_{\alpha'} \rangle$ expectation values,
cf.\ Eq.~\eqref{eq:SigmaIFG_weight_rephrased},
while the corresponding results from $\Sigma^\FG$ deviate by roughly 15\% in each component.
The insets, enlarging the low-energy regime, reveal that, also in the non-diagonal setting,
$\Sigma^\IFG$ is free from the wiggles present in $\Sigma^\FG$.
Remarkably, this applies not only to the diagonal self-energy component where the imaginary part vanishes at $\nu=0$,
but also to the off-diagonal component where this value is finite.

\section{Anderson impurity model at strong interaction}
\label{sec:appendix2}

Here, we give additional numerical results for the ($\alpha$-diagonal) single-orbital Anderson impurity model at strong interaction.
We choose the same parameter set as in Ref.~\cite{Bulla1998}:
a box-shaped hybridization function of half bandwidth $D=1$ and
strength $\Gamma=0.015$,
an interaction value of $U=0.2$,
and $\epsilon_d = -0.1$ corresponding to half filling.
The particle-hole symmetry allows us to exploit 
SU(2) charge and SU(2) spin symmetry in the calculation, as already done in Sec.~\ref{sec:Hubbard}.
We set $\Lambda=2$ and $n_z=4$ as in Sec.~\ref{sec:SIAM}.
The temperature $T=10^{-8}$ is again far below 
the Kondo temperature of
$T_\mK \approx 2.06 \cdot 10^{-4}$,
following from the same formula as used in Sec.~\ref{sec:SIAM}.

Figure~\ref{fig:SIAM_SE_largeU} is analogous to Fig.~\ref{fig:SIAM_SE};
we restrict panels (a) and (b) to positive frequencies in light of particle-hole symmetry.
The findings from Sec.~\ref{sec:SIAM} also hold analogously in the current setting at strong interaction:
$\Sigma^\FG$ and $\Sigma^\IFG$ are overall consistent;
deviations at large frequencies in Figs.~\ref{fig:SIAM_SE_largeU}(a) and \ref{fig:SIAM_SE_largeU}(b)
reflect the fact that the high-frequency asymptote in the real part
and the total weight in the imaginary part 
are exactly fulfilled by $\Sigma^\IFG_\nu$,
whereas this is not the case for  $\Sigma^\FG_\nu$.
The agreement between $\Re\Sigma^\IFG_\nu$ and $\Sigma^{(0)}_\ex/\nu$ for large $\nu$ 
can be seen in the inset of Fig.~\ref{fig:SIAM_SE_largeU}(a).

The low-energy behavior of $\Im \Sigma_\nu$ for
an increasing number of kept states $N_\kp$ [SU(2)$\times$SU(2) multiplets]
is compared in Fig.~\ref{fig:SIAM_SE_largeU}(c).
The wiggles of $\Im \Sigma^\FG_\nu$ at low energies,
as for instance observed in Fig.~\ref{fig:SIAM_SE}(c),
are absent in this particle-hole symmetric setting.
However, for $N_\kp =1000$, 
$-\Im \Sigma^\FG_\nu$ still overshoots to negative values on the scale of $10^{-3}\Gamma$.
Increasing $N_\kp$ to $2000$ and $5000$,
$-\Im \Sigma^\FG_\nu$ continues to shift:
it comes closer to the $\nu$ axis without fully reaching it,
violating $\Im \Sigma^\FG(0)=0$ with errors on the order of $10^{-3}\Gamma$.
By contrast, $-\Im \Sigma^\IFG_\nu$ shows a clean, nonnegative parabola which,
on the given scale,
has its vertex right at the origin
and is converged for $N_\kp$ as low as $500$.

\bibliographystyle{apsrev4-1}
\bibliography{references}

\begin{thebibliography}{54}%
\makeatletter
\providecommand \@ifxundefined [1]{%
 \@ifx{#1\undefined}
}%
\providecommand \@ifnum [1]{%
 \ifnum #1\expandafter \@firstoftwo
 \else \expandafter \@secondoftwo
 \fi
}%
\providecommand \@ifx [1]{%
 \ifx #1\expandafter \@firstoftwo
 \else \expandafter \@secondoftwo
 \fi
}%
\providecommand \natexlab [1]{#1}%
\providecommand \enquote  [1]{``#1''}%
\providecommand \bibnamefont  [1]{#1}%
\providecommand \bibfnamefont [1]{#1}%
\providecommand \citenamefont [1]{#1}%
\providecommand \href@noop [0]{\@secondoftwo}%
\providecommand \href [0]{\begingroup \@sanitize@url \@href}%
\providecommand \@href[1]{\@@startlink{#1}\@@href}%
\providecommand \@@href[1]{\endgroup#1\@@endlink}%
\providecommand \@sanitize@url [0]{\catcode `\\12\catcode `\$12\catcode
  `\&12\catcode `\#12\catcode `\^12\catcode `\_12\catcode `\%12\relax}%
\providecommand \@@startlink[1]{}%
\providecommand \@@endlink[0]{}%
\providecommand \url  [0]{\begingroup\@sanitize@url \@url }%
\providecommand \@url [1]{\endgroup\@href {#1}{\urlprefix }}%
\providecommand \urlprefix  [0]{URL }%
\providecommand \Eprint [0]{\href }%
\providecommand \doibase [0]{http://dx.doi.org/}%
\providecommand \selectlanguage [0]{\@gobble}%
\providecommand \bibinfo  [0]{\@secondoftwo}%
\providecommand \bibfield  [0]{\@secondoftwo}%
\providecommand \translation [1]{[#1]}%
\providecommand \BibitemOpen [0]{}%
\providecommand \bibitemStop [0]{}%
\providecommand \bibitemNoStop [0]{.\EOS\space}%
\providecommand \EOS [0]{\spacefactor3000\relax}%
\providecommand \BibitemShut  [1]{\csname bibitem#1\endcsname}%
\let\auto@bib@innerbib\@empty
\bibitem [{\citenamefont {Hanson}\ \emph {et~al.}(2007)\citenamefont {Hanson},
  \citenamefont {Kouwenhoven}, \citenamefont {Petta}, \citenamefont {Tarucha},\
  and\ \citenamefont {Vandersypen}}]{Hanson2007}%
  \BibitemOpen
  \bibfield  {author} {\bibinfo {author} {\bibfnamefont {R.}~\bibnamefont
  {Hanson}}, \bibinfo {author} {\bibfnamefont {L.~P.}\ \bibnamefont
  {Kouwenhoven}}, \bibinfo {author} {\bibfnamefont {J.~R.}\ \bibnamefont
  {Petta}}, \bibinfo {author} {\bibfnamefont {S.}~\bibnamefont {Tarucha}}, \
  and\ \bibinfo {author} {\bibfnamefont {L.~M.~K.}\ \bibnamefont
  {Vandersypen}},\ }\href {\doibase 10.1103/RevModPhys.79.1217} {\bibfield
  {journal} {\bibinfo  {journal} {Rev. Mod. Phys.}\ }\textbf {\bibinfo {volume}
  {79}},\ \bibinfo {pages} {1217} (\bibinfo {year} {2007})}\BibitemShut
  {NoStop}%
\bibitem [{\citenamefont {Georges}\ \emph {et~al.}(1996)\citenamefont
  {Georges}, \citenamefont {Kotliar}, \citenamefont {Krauth},\ and\
  \citenamefont {Rozenberg}}]{Georges1996}%
  \BibitemOpen
  \bibfield  {author} {\bibinfo {author} {\bibfnamefont {A.}~\bibnamefont
  {Georges}}, \bibinfo {author} {\bibfnamefont {G.}~\bibnamefont {Kotliar}},
  \bibinfo {author} {\bibfnamefont {W.}~\bibnamefont {Krauth}}, \ and\ \bibinfo
  {author} {\bibfnamefont {M.~J.}\ \bibnamefont {Rozenberg}},\ }\href {\doibase
  10.1103/RevModPhys.68.13} {\bibfield  {journal} {\bibinfo  {journal} {Rev.
  Mod. Phys.}\ }\textbf {\bibinfo {volume} {68}},\ \bibinfo {pages} {13}
  (\bibinfo {year} {1996})}\BibitemShut {NoStop}%
\bibitem [{\citenamefont {Wilson}(1975)}]{Wilson1975}%
  \BibitemOpen
  \bibfield  {author} {\bibinfo {author} {\bibfnamefont {K.~G.}\ \bibnamefont
  {Wilson}},\ }\href {\doibase 10.1103/RevModPhys.47.773} {\bibfield  {journal}
  {\bibinfo  {journal} {Rev. Mod. Phys.}\ }\textbf {\bibinfo {volume} {47}},\
  \bibinfo {pages} {773} (\bibinfo {year} {1975})}\BibitemShut {NoStop}%
\bibitem [{\citenamefont {Bulla}\ \emph {et~al.}(2008)\citenamefont {Bulla},
  \citenamefont {Costi},\ and\ \citenamefont {Pruschke}}]{Bulla2008}%
  \BibitemOpen
  \bibfield  {author} {\bibinfo {author} {\bibfnamefont {R.}~\bibnamefont
  {Bulla}}, \bibinfo {author} {\bibfnamefont {T.~A.}\ \bibnamefont {Costi}}, \
  and\ \bibinfo {author} {\bibfnamefont {T.}~\bibnamefont {Pruschke}},\ }\href
  {\doibase 10.1103/RevModPhys.80.395} {\bibfield  {journal} {\bibinfo
  {journal} {Rev. Mod. Phys.}\ }\textbf {\bibinfo {volume} {80}},\ \bibinfo
  {pages} {395} (\bibinfo {year} {2008})}\BibitemShut {NoStop}%
\bibitem [{\citenamefont {Bulla}(1999)}]{Bulla1999}%
  \BibitemOpen
  \bibfield  {author} {\bibinfo {author} {\bibfnamefont {R.}~\bibnamefont
  {Bulla}},\ }\href {\doibase 10.1103/PhysRevLett.83.136} {\bibfield  {journal}
  {\bibinfo  {journal} {Phys. Rev. Lett.}\ }\textbf {\bibinfo {volume} {83}},\
  \bibinfo {pages} {136} (\bibinfo {year} {1999})}\BibitemShut {NoStop}%
\bibitem [{\citenamefont {Bulla}\ \emph {et~al.}(2001)\citenamefont {Bulla},
  \citenamefont {Costi},\ and\ \citenamefont {Vollhardt}}]{Bulla2001}%
  \BibitemOpen
  \bibfield  {author} {\bibinfo {author} {\bibfnamefont {R.}~\bibnamefont
  {Bulla}}, \bibinfo {author} {\bibfnamefont {T.~A.}\ \bibnamefont {Costi}}, \
  and\ \bibinfo {author} {\bibfnamefont {D.}~\bibnamefont {Vollhardt}},\ }\href
  {\doibase 10.1103/PhysRevB.64.045103} {\bibfield  {journal} {\bibinfo
  {journal} {Phys. Rev. B}\ }\textbf {\bibinfo {volume} {64}},\ \bibinfo
  {pages} {045103} (\bibinfo {year} {2001})}\BibitemShut {NoStop}%
\bibitem [{\citenamefont {Deng}\ \emph {et~al.}(2013)\citenamefont {Deng},
  \citenamefont {Mravlje}, \citenamefont {\ifmmode~\check{Z}\else
  \v{Z}\fi{}itko}, \citenamefont {Ferrero}, \citenamefont {Kotliar},\ and\
  \citenamefont {Georges}}]{Deng2013}%
  \BibitemOpen
  \bibfield  {author} {\bibinfo {author} {\bibfnamefont {X.}~\bibnamefont
  {Deng}}, \bibinfo {author} {\bibfnamefont {J.}~\bibnamefont {Mravlje}},
  \bibinfo {author} {\bibfnamefont {R.}~\bibnamefont {\ifmmode~\check{Z}\else
  \v{Z}\fi{}itko}}, \bibinfo {author} {\bibfnamefont {M.}~\bibnamefont
  {Ferrero}}, \bibinfo {author} {\bibfnamefont {G.}~\bibnamefont {Kotliar}}, \
  and\ \bibinfo {author} {\bibfnamefont {A.}~\bibnamefont {Georges}},\ }\href
  {\doibase 10.1103/PhysRevLett.110.086401} {\bibfield  {journal} {\bibinfo
  {journal} {Phys. Rev. Lett.}\ }\textbf {\bibinfo {volume} {110}},\ \bibinfo
  {pages} {086401} (\bibinfo {year} {2013})}\BibitemShut {NoStop}%
\bibitem [{\citenamefont {Lee}\ \emph {et~al.}(2017)\citenamefont {Lee},
  \citenamefont {von Delft},\ and\ \citenamefont {Weichselbaum}}]{Lee2017}%
  \BibitemOpen
  \bibfield  {author} {\bibinfo {author} {\bibfnamefont {S.-S.~B.}\
  \bibnamefont {Lee}}, \bibinfo {author} {\bibfnamefont {J.}~\bibnamefont {von
  Delft}}, \ and\ \bibinfo {author} {\bibfnamefont {A.}~\bibnamefont
  {Weichselbaum}},\ }\href {\doibase 10.1103/PhysRevLett.119.236402} {\bibfield
   {journal} {\bibinfo  {journal} {Phys. Rev. Lett.}\ }\textbf {\bibinfo
  {volume} {119}},\ \bibinfo {pages} {236402} (\bibinfo {year}
  {2017})}\BibitemShut {NoStop}%
\bibitem [{\citenamefont {Vu\ifmmode \check{c}\else \v{c}\fi{}i\ifmmode
  \check{c}\else \v{c}\fi{}evi\ifmmode~\acute{c}\else \'{c}\fi{}}\ \emph
  {et~al.}(2019)\citenamefont {Vu\ifmmode \check{c}\else \v{c}\fi{}i\ifmmode
  \check{c}\else \v{c}\fi{}evi\ifmmode~\acute{c}\else \'{c}\fi{}},
  \citenamefont {Kokalj}, \citenamefont {\ifmmode~\check{Z}\else
  \v{Z}\fi{}itko}, \citenamefont {Wentzell}, \citenamefont
  {Tanaskovi\ifmmode~\acute{c}\else \'{c}\fi{}},\ and\ \citenamefont
  {Mravlje}}]{Vucicevic2019}%
  \BibitemOpen
  \bibfield  {author} {\bibinfo {author} {\bibfnamefont {J.}~\bibnamefont
  {Vu\ifmmode \check{c}\else \v{c}\fi{}i\ifmmode \check{c}\else
  \v{c}\fi{}evi\ifmmode~\acute{c}\else \'{c}\fi{}}}, \bibinfo {author}
  {\bibfnamefont {J.}~\bibnamefont {Kokalj}}, \bibinfo {author} {\bibfnamefont
  {R.}~\bibnamefont {\ifmmode~\check{Z}\else \v{Z}\fi{}itko}}, \bibinfo
  {author} {\bibfnamefont {N.}~\bibnamefont {Wentzell}}, \bibinfo {author}
  {\bibfnamefont {D.}~\bibnamefont {Tanaskovi\ifmmode~\acute{c}\else
  \'{c}\fi{}}}, \ and\ \bibinfo {author} {\bibfnamefont {J.}~\bibnamefont
  {Mravlje}},\ }\href {\doibase 10.1103/PhysRevLett.123.036601} {\bibfield
  {journal} {\bibinfo  {journal} {Phys. Rev. Lett.}\ }\textbf {\bibinfo
  {volume} {123}},\ \bibinfo {pages} {036601} (\bibinfo {year}
  {2019})}\BibitemShut {NoStop}%
\bibitem [{\citenamefont {Vrani\ifmmode~\acute{c}\else \'{c}\fi{}}\ \emph
  {et~al.}(2020)\citenamefont {Vrani\ifmmode~\acute{c}\else \'{c}\fi{}},
  \citenamefont {Vu\ifmmode \check{c}\else \v{c}\fi{}i\ifmmode \check{c}\else
  \v{c}\fi{}evi\ifmmode~\acute{c}\else \'{c}\fi{}}, \citenamefont {Kokalj},
  \citenamefont {Skolimowski}, \citenamefont {\ifmmode~\check{Z}\else
  \v{Z}\fi{}itko}, \citenamefont {Mravlje},\ and\ \citenamefont
  {Tanaskovi\ifmmode~\acute{c}\else \'{c}\fi{}}}]{Vranic2020}%
  \BibitemOpen
  \bibfield  {author} {\bibinfo {author} {\bibfnamefont {A.}~\bibnamefont
  {Vrani\ifmmode~\acute{c}\else \'{c}\fi{}}}, \bibinfo {author} {\bibfnamefont
  {J.}~\bibnamefont {Vu\ifmmode \check{c}\else \v{c}\fi{}i\ifmmode
  \check{c}\else \v{c}\fi{}evi\ifmmode~\acute{c}\else \'{c}\fi{}}}, \bibinfo
  {author} {\bibfnamefont {J.}~\bibnamefont {Kokalj}}, \bibinfo {author}
  {\bibfnamefont {J.}~\bibnamefont {Skolimowski}}, \bibinfo {author}
  {\bibfnamefont {R.}~\bibnamefont {\ifmmode~\check{Z}\else \v{Z}\fi{}itko}},
  \bibinfo {author} {\bibfnamefont {J.}~\bibnamefont {Mravlje}}, \ and\
  \bibinfo {author} {\bibfnamefont {D.}~\bibnamefont
  {Tanaskovi\ifmmode~\acute{c}\else \'{c}\fi{}}},\ }\href {\doibase
  10.1103/PhysRevB.102.115142} {\bibfield  {journal} {\bibinfo  {journal}
  {Phys. Rev. B}\ }\textbf {\bibinfo {volume} {102}},\ \bibinfo {pages}
  {115142} (\bibinfo {year} {2020})}\BibitemShut {NoStop}%
\bibitem [{\citenamefont {Vu\ifmmode \check{c}\else \v{c}\fi{}i\ifmmode
  \check{c}\else \v{c}\fi{}evi\ifmmode~\acute{c}\else \'{c}\fi{}}\ and\
  \citenamefont {\ifmmode~\check{Z}\else
  \v{Z}\fi{}itko}(2021{\natexlab{a}})}]{Vucicevic2021a}%
  \BibitemOpen
  \bibfield  {author} {\bibinfo {author} {\bibfnamefont {J.}~\bibnamefont
  {Vu\ifmmode \check{c}\else \v{c}\fi{}i\ifmmode \check{c}\else
  \v{c}\fi{}evi\ifmmode~\acute{c}\else \'{c}\fi{}}}\ and\ \bibinfo {author}
  {\bibfnamefont {R.}~\bibnamefont {\ifmmode~\check{Z}\else \v{Z}\fi{}itko}},\
  }\href {\doibase 10.1103/PhysRevLett.127.196601} {\bibfield  {journal}
  {\bibinfo  {journal} {Phys. Rev. Lett.}\ }\textbf {\bibinfo {volume} {127}},\
  \bibinfo {pages} {196601} (\bibinfo {year} {2021}{\natexlab{a}})}\BibitemShut
  {NoStop}%
\bibitem [{\citenamefont {Vu\ifmmode \check{c}\else \v{c}\fi{}i\ifmmode
  \check{c}\else \v{c}\fi{}evi\ifmmode~\acute{c}\else \'{c}\fi{}}\ and\
  \citenamefont {\ifmmode~\check{Z}\else
  \v{Z}\fi{}itko}(2021{\natexlab{b}})}]{Vucicevic2021b}%
  \BibitemOpen
  \bibfield  {author} {\bibinfo {author} {\bibfnamefont {J.}~\bibnamefont
  {Vu\ifmmode \check{c}\else \v{c}\fi{}i\ifmmode \check{c}\else
  \v{c}\fi{}evi\ifmmode~\acute{c}\else \'{c}\fi{}}}\ and\ \bibinfo {author}
  {\bibfnamefont {R.}~\bibnamefont {\ifmmode~\check{Z}\else \v{Z}\fi{}itko}},\
  }\href {\doibase 10.1103/PhysRevB.104.205101} {\bibfield  {journal} {\bibinfo
   {journal} {Phys. Rev. B}\ }\textbf {\bibinfo {volume} {104}},\ \bibinfo
  {pages} {205101} (\bibinfo {year} {2021}{\natexlab{b}})}\BibitemShut
  {NoStop}%
\bibitem [{\citenamefont {Pruschke}\ and\ \citenamefont
  {Bulla}(2005)}]{Pruschke2005}%
  \BibitemOpen
  \bibfield  {author} {\bibinfo {author} {\bibfnamefont {T.}~\bibnamefont
  {Pruschke}}\ and\ \bibinfo {author} {\bibfnamefont {R.}~\bibnamefont
  {Bulla}},\ }\href {\doibase 10.1140/epjb/e2005-00117-4} {\bibfield  {journal}
  {\bibinfo  {journal} {Eur. Phys. J. B}\ }\textbf {\bibinfo {volume} {44}},\
  \bibinfo {pages} {217} (\bibinfo {year} {2005})}\BibitemShut {NoStop}%
\bibitem [{\citenamefont {Peters}\ and\ \citenamefont
  {Pruschke}(2010{\natexlab{a}})}]{Peters2010a}%
  \BibitemOpen
  \bibfield  {author} {\bibinfo {author} {\bibfnamefont {R.}~\bibnamefont
  {Peters}}\ and\ \bibinfo {author} {\bibfnamefont {T.}~\bibnamefont
  {Pruschke}},\ }\href {\doibase 10.1103/PhysRevB.81.035112} {\bibfield
  {journal} {\bibinfo  {journal} {Phys. Rev. B}\ }\textbf {\bibinfo {volume}
  {81}},\ \bibinfo {pages} {035112} (\bibinfo {year}
  {2010}{\natexlab{a}})}\BibitemShut {NoStop}%
\bibitem [{\citenamefont {Peters}\ and\ \citenamefont
  {Pruschke}(2010{\natexlab{b}})}]{Peters2010b}%
  \BibitemOpen
  \bibfield  {author} {\bibinfo {author} {\bibfnamefont {R.}~\bibnamefont
  {Peters}}\ and\ \bibinfo {author} {\bibfnamefont {T.}~\bibnamefont
  {Pruschke}},\ }\href {\doibase 10.1088/1742-6596/200/1/012158} {\bibfield
  {journal} {\bibinfo  {journal} {J. Phys. Conf. Ser.}\ }\textbf {\bibinfo
  {volume} {200}},\ \bibinfo {pages} {012158} (\bibinfo {year}
  {2010}{\natexlab{b}})}\BibitemShut {NoStop}%
\bibitem [{\citenamefont {Peters}\ \emph {et~al.}(2011)\citenamefont {Peters},
  \citenamefont {Kawakami},\ and\ \citenamefont {Pruschke}}]{Peters2011}%
  \BibitemOpen
  \bibfield  {author} {\bibinfo {author} {\bibfnamefont {R.}~\bibnamefont
  {Peters}}, \bibinfo {author} {\bibfnamefont {N.}~\bibnamefont {Kawakami}}, \
  and\ \bibinfo {author} {\bibfnamefont {T.}~\bibnamefont {Pruschke}},\ }\href
  {\doibase 10.1103/PhysRevB.83.125110} {\bibfield  {journal} {\bibinfo
  {journal} {Phys. Rev. B}\ }\textbf {\bibinfo {volume} {83}},\ \bibinfo
  {pages} {125110} (\bibinfo {year} {2011})}\BibitemShut {NoStop}%
\bibitem [{\citenamefont {Greger}\ \emph {et~al.}(2013)\citenamefont {Greger},
  \citenamefont {Kollar},\ and\ \citenamefont {Vollhardt}}]{Greger2013a}%
  \BibitemOpen
  \bibfield  {author} {\bibinfo {author} {\bibfnamefont {M.}~\bibnamefont
  {Greger}}, \bibinfo {author} {\bibfnamefont {M.}~\bibnamefont {Kollar}}, \
  and\ \bibinfo {author} {\bibfnamefont {D.}~\bibnamefont {Vollhardt}},\ }\href
  {\doibase 10.1103/PhysRevLett.110.046403} {\bibfield  {journal} {\bibinfo
  {journal} {Phys. Rev. Lett.}\ }\textbf {\bibinfo {volume} {110}},\ \bibinfo
  {pages} {046403} (\bibinfo {year} {2013})}\BibitemShut {NoStop}%
\bibitem [{\citenamefont {Greger}\ \emph {et~al.}()\citenamefont {Greger},
  \citenamefont {Sekania},\ and\ \citenamefont {Kollar}}]{Greger2013b}%
  \BibitemOpen
  \bibfield  {author} {\bibinfo {author} {\bibfnamefont {M.}~\bibnamefont
  {Greger}}, \bibinfo {author} {\bibfnamefont {M.}~\bibnamefont {Sekania}}, \
  and\ \bibinfo {author} {\bibfnamefont {M.}~\bibnamefont {Kollar}},\ }\href
  {http://arxiv.org/abs/1312.0100} {\ }\Eprint {http://arxiv.org/abs/1312.0100}
  {arXiv:1312.0100} \BibitemShut {NoStop}%
\bibitem [{\citenamefont {Kugler}\ and\ \citenamefont
  {Kotliar}()}]{Kugler2021}%
  \BibitemOpen
  \bibfield  {author} {\bibinfo {author} {\bibfnamefont {F.~B.}\ \bibnamefont
  {Kugler}}\ and\ \bibinfo {author} {\bibfnamefont {G.}~\bibnamefont
  {Kotliar}},\ }\href {https://arxiv.org/abs/2112.14691} {\ }\Eprint
  {http://arxiv.org/abs/2112.14691} {arXiv:2112.14691} \BibitemShut {NoStop}%
\bibitem [{\citenamefont {Stadler}\ \emph {et~al.}(2016)\citenamefont
  {Stadler}, \citenamefont {Mitchell}, \citenamefont {von Delft},\ and\
  \citenamefont {Weichselbaum}}]{Stadler2016}%
  \BibitemOpen
  \bibfield  {author} {\bibinfo {author} {\bibfnamefont {K.~M.}\ \bibnamefont
  {Stadler}}, \bibinfo {author} {\bibfnamefont {A.~K.}\ \bibnamefont
  {Mitchell}}, \bibinfo {author} {\bibfnamefont {J.}~\bibnamefont {von Delft}},
  \ and\ \bibinfo {author} {\bibfnamefont {A.}~\bibnamefont {Weichselbaum}},\
  }\href {\doibase 10.1103/PhysRevB.93.235101} {\bibfield  {journal} {\bibinfo
  {journal} {Phys. Rev. B}\ }\textbf {\bibinfo {volume} {93}},\ \bibinfo
  {pages} {235101} (\bibinfo {year} {2016})}\BibitemShut {NoStop}%
\bibitem [{\citenamefont {Stadler}\ \emph {et~al.}(2019)\citenamefont
  {Stadler}, \citenamefont {Kotliar}, \citenamefont {Weichselbaum},\ and\
  \citenamefont {von Delft}}]{Stadler2019}%
  \BibitemOpen
  \bibfield  {author} {\bibinfo {author} {\bibfnamefont {K.~M.}\ \bibnamefont
  {Stadler}}, \bibinfo {author} {\bibfnamefont {G.}~\bibnamefont {Kotliar}},
  \bibinfo {author} {\bibfnamefont {A.}~\bibnamefont {Weichselbaum}}, \ and\
  \bibinfo {author} {\bibfnamefont {J.}~\bibnamefont {von Delft}},\ }\href
  {\doibase 10.1016/j.aop.2018.10.017} {\bibfield  {journal} {\bibinfo
  {journal} {Ann. Phys.}\ }\textbf {\bibinfo {volume} {405}},\ \bibinfo {pages}
  {365 } (\bibinfo {year} {2019})}\BibitemShut {NoStop}%
\bibitem [{\citenamefont {Kugler}\ \emph {et~al.}(2019)\citenamefont {Kugler},
  \citenamefont {Lee}, \citenamefont {Weichselbaum}, \citenamefont {Kotliar},\
  and\ \citenamefont {von Delft}}]{Kugler2019}%
  \BibitemOpen
  \bibfield  {author} {\bibinfo {author} {\bibfnamefont {F.~B.}\ \bibnamefont
  {Kugler}}, \bibinfo {author} {\bibfnamefont {S.-S.~B.}\ \bibnamefont {Lee}},
  \bibinfo {author} {\bibfnamefont {A.}~\bibnamefont {Weichselbaum}}, \bibinfo
  {author} {\bibfnamefont {G.}~\bibnamefont {Kotliar}}, \ and\ \bibinfo
  {author} {\bibfnamefont {J.}~\bibnamefont {von Delft}},\ }\href {\doibase
  10.1103/PhysRevB.100.115159} {\bibfield  {journal} {\bibinfo  {journal}
  {Phys. Rev. B}\ }\textbf {\bibinfo {volume} {100}},\ \bibinfo {pages}
  {115159} (\bibinfo {year} {2019})}\BibitemShut {NoStop}%
\bibitem [{\citenamefont {Stadler}\ \emph {et~al.}(2021)\citenamefont
  {Stadler}, \citenamefont {Kotliar}, \citenamefont {Lee}, \citenamefont
  {Weichselbaum},\ and\ \citenamefont {von Delft}}]{Stadler2021}%
  \BibitemOpen
  \bibfield  {author} {\bibinfo {author} {\bibfnamefont {K.~M.}\ \bibnamefont
  {Stadler}}, \bibinfo {author} {\bibfnamefont {G.}~\bibnamefont {Kotliar}},
  \bibinfo {author} {\bibfnamefont {S.-S.~B.}\ \bibnamefont {Lee}}, \bibinfo
  {author} {\bibfnamefont {A.}~\bibnamefont {Weichselbaum}}, \ and\ \bibinfo
  {author} {\bibfnamefont {J.}~\bibnamefont {von Delft}},\ }\href {\doibase
  10.1103/PhysRevB.104.115107} {\bibfield  {journal} {\bibinfo  {journal}
  {Phys. Rev. B}\ }\textbf {\bibinfo {volume} {104}},\ \bibinfo {pages}
  {115107} (\bibinfo {year} {2021})}\BibitemShut {NoStop}%
\bibitem [{\citenamefont {Kugler}\ \emph {et~al.}(2020)\citenamefont {Kugler},
  \citenamefont {Zingl}, \citenamefont {Strand}, \citenamefont {Lee},
  \citenamefont {von Delft},\ and\ \citenamefont {Georges}}]{Kugler2020}%
  \BibitemOpen
  \bibfield  {author} {\bibinfo {author} {\bibfnamefont {F.~B.}\ \bibnamefont
  {Kugler}}, \bibinfo {author} {\bibfnamefont {M.}~\bibnamefont {Zingl}},
  \bibinfo {author} {\bibfnamefont {H.~U.~R.}\ \bibnamefont {Strand}}, \bibinfo
  {author} {\bibfnamefont {S.-S.~B.}\ \bibnamefont {Lee}}, \bibinfo {author}
  {\bibfnamefont {J.}~\bibnamefont {von Delft}}, \ and\ \bibinfo {author}
  {\bibfnamefont {A.}~\bibnamefont {Georges}},\ }\href {\doibase
  10.1103/PhysRevLett.124.016401} {\bibfield  {journal} {\bibinfo  {journal}
  {Phys. Rev. Lett.}\ }\textbf {\bibinfo {volume} {124}},\ \bibinfo {pages}
  {016401} (\bibinfo {year} {2020})}\BibitemShut {NoStop}%
\bibitem [{\citenamefont {Peters}\ \emph {et~al.}(2006)\citenamefont {Peters},
  \citenamefont {Pruschke},\ and\ \citenamefont {Anders}}]{Peters2006}%
  \BibitemOpen
  \bibfield  {author} {\bibinfo {author} {\bibfnamefont {R.}~\bibnamefont
  {Peters}}, \bibinfo {author} {\bibfnamefont {T.}~\bibnamefont {Pruschke}}, \
  and\ \bibinfo {author} {\bibfnamefont {F.~B.}\ \bibnamefont {Anders}},\
  }\href {\doibase 10.1103/PhysRevB.74.245114} {\bibfield  {journal} {\bibinfo
  {journal} {Phys. Rev. B}\ }\textbf {\bibinfo {volume} {74}},\ \bibinfo
  {pages} {245114} (\bibinfo {year} {2006})}\BibitemShut {NoStop}%
\bibitem [{\citenamefont {Weichselbaum}\ and\ \citenamefont {von
  Delft}(2007)}]{Weichselbaum2007}%
  \BibitemOpen
  \bibfield  {author} {\bibinfo {author} {\bibfnamefont {A.}~\bibnamefont
  {Weichselbaum}}\ and\ \bibinfo {author} {\bibfnamefont {J.}~\bibnamefont {von
  Delft}},\ }\href {\doibase 10.1103/PhysRevLett.99.076402} {\bibfield
  {journal} {\bibinfo  {journal} {Phys. Rev. Lett.}\ }\textbf {\bibinfo
  {volume} {99}},\ \bibinfo {pages} {076402} (\bibinfo {year}
  {2007})}\BibitemShut {NoStop}%
\bibitem [{\citenamefont {Bulla}\ \emph {et~al.}(1998)\citenamefont {Bulla},
  \citenamefont {Hewson},\ and\ \citenamefont {Pruschke}}]{Bulla1998}%
  \BibitemOpen
  \bibfield  {author} {\bibinfo {author} {\bibfnamefont {R.}~\bibnamefont
  {Bulla}}, \bibinfo {author} {\bibfnamefont {A.~C.}\ \bibnamefont {Hewson}}, \
  and\ \bibinfo {author} {\bibfnamefont {T.}~\bibnamefont {Pruschke}},\ }\href
  {\doibase 10.1088/0953-8984/10/37/021} {\bibfield  {journal} {\bibinfo
  {journal} {J. Phys.: Condens. Matter}\ }\textbf {\bibinfo {volume} {10}},\
  \bibinfo {pages} {8365} (\bibinfo {year} {1998})}\BibitemShut {NoStop}%
\bibitem [{\citenamefont {\ifmmode~\check{Z}\else \v{Z}\fi{}itko}\ and\
  \citenamefont {Pruschke}(2009)}]{Zitko2009}%
  \BibitemOpen
  \bibfield  {author} {\bibinfo {author} {\bibfnamefont {R.}~\bibnamefont
  {\ifmmode~\check{Z}\else \v{Z}\fi{}itko}}\ and\ \bibinfo {author}
  {\bibfnamefont {T.}~\bibnamefont {Pruschke}},\ }\href {\doibase
  10.1103/PhysRevB.79.085106} {\bibfield  {journal} {\bibinfo  {journal} {Phys.
  Rev. B}\ }\textbf {\bibinfo {volume} {79}},\ \bibinfo {pages} {085106}
  (\bibinfo {year} {2009})}\BibitemShut {NoStop}%
\bibitem [{\citenamefont {Lee}\ and\ \citenamefont
  {Weichselbaum}(2016)}]{Lee2016}%
  \BibitemOpen
  \bibfield  {author} {\bibinfo {author} {\bibfnamefont {S.-S.~B.}\
  \bibnamefont {Lee}}\ and\ \bibinfo {author} {\bibfnamefont {A.}~\bibnamefont
  {Weichselbaum}},\ }\href {\doibase 10.1103/PhysRevB.94.235127} {\bibfield
  {journal} {\bibinfo  {journal} {Phys. Rev. B}\ }\textbf {\bibinfo {volume}
  {94}},\ \bibinfo {pages} {235127} (\bibinfo {year} {2016})}\BibitemShut
  {NoStop}%
\bibitem [{\citenamefont {Kaufmann}\ \emph {et~al.}(2019)\citenamefont
  {Kaufmann}, \citenamefont {Gunacker}, \citenamefont {Kowalski}, \citenamefont
  {Sangiovanni},\ and\ \citenamefont {Held}}]{Kaufmann2019}%
  \BibitemOpen
  \bibfield  {author} {\bibinfo {author} {\bibfnamefont {J.}~\bibnamefont
  {Kaufmann}}, \bibinfo {author} {\bibfnamefont {P.}~\bibnamefont {Gunacker}},
  \bibinfo {author} {\bibfnamefont {A.}~\bibnamefont {Kowalski}}, \bibinfo
  {author} {\bibfnamefont {G.}~\bibnamefont {Sangiovanni}}, \ and\ \bibinfo
  {author} {\bibfnamefont {K.}~\bibnamefont {Held}},\ }\href {\doibase
  10.1103/PhysRevB.100.075119} {\bibfield  {journal} {\bibinfo  {journal}
  {Phys. Rev. B}\ }\textbf {\bibinfo {volume} {100}},\ \bibinfo {pages}
  {075119} (\bibinfo {year} {2019})}\BibitemShut {NoStop}%
\bibitem [{\citenamefont {Anderson}(1961)}]{Anderson1961}%
  \BibitemOpen
  \bibfield  {author} {\bibinfo {author} {\bibfnamefont {P.~W.}\ \bibnamefont
  {Anderson}},\ }\href {\doibase 10.1103/PhysRev.124.41} {\bibfield  {journal}
  {\bibinfo  {journal} {Phys. Rev.}\ }\textbf {\bibinfo {volume} {124}},\
  \bibinfo {pages} {41} (\bibinfo {year} {1961})}\BibitemShut {NoStop}%
\bibitem [{\citenamefont {Dworin}\ and\ \citenamefont
  {Narath}(1970)}]{Dworin1970}%
  \BibitemOpen
  \bibfield  {author} {\bibinfo {author} {\bibfnamefont {L.}~\bibnamefont
  {Dworin}}\ and\ \bibinfo {author} {\bibfnamefont {A.}~\bibnamefont
  {Narath}},\ }\href {\doibase 10.1103/PhysRevLett.25.1287} {\bibfield
  {journal} {\bibinfo  {journal} {Phys. Rev. Lett.}\ }\textbf {\bibinfo
  {volume} {25}},\ \bibinfo {pages} {1287} (\bibinfo {year}
  {1970})}\BibitemShut {NoStop}%
\bibitem [{\citenamefont {Georges}\ \emph {et~al.}(2013)\citenamefont
  {Georges}, \citenamefont {de' Medici},\ and\ \citenamefont
  {Mravlje}}]{Georges2013}%
  \BibitemOpen
  \bibfield  {author} {\bibinfo {author} {\bibfnamefont {A.}~\bibnamefont
  {Georges}}, \bibinfo {author} {\bibfnamefont {L.}~\bibnamefont {de' Medici}},
  \ and\ \bibinfo {author} {\bibfnamefont {J.}~\bibnamefont {Mravlje}},\ }\href
  {\doibase 10.1146/annurev-conmatphys-020911-125045} {\bibfield  {journal}
  {\bibinfo  {journal} {Annu. Rev. Condens. Matter Phys.}\ }\textbf {\bibinfo
  {volume} {4}},\ \bibinfo {pages} {137} (\bibinfo {year} {2013})}\BibitemShut
  {NoStop}%
\bibitem [{\citenamefont {Hafermann}\ \emph {et~al.}(2012)\citenamefont
  {Hafermann}, \citenamefont {Patton},\ and\ \citenamefont
  {Werner}}]{Hafermann2012}%
  \BibitemOpen
  \bibfield  {author} {\bibinfo {author} {\bibfnamefont {H.}~\bibnamefont
  {Hafermann}}, \bibinfo {author} {\bibfnamefont {K.~R.}\ \bibnamefont
  {Patton}}, \ and\ \bibinfo {author} {\bibfnamefont {P.}~\bibnamefont
  {Werner}},\ }\href {\doibase 10.1103/PhysRevB.85.205106} {\bibfield
  {journal} {\bibinfo  {journal} {Phys. Rev. B}\ }\textbf {\bibinfo {volume}
  {85}},\ \bibinfo {pages} {205106} (\bibinfo {year} {2012})}\BibitemShut
  {NoStop}%
\bibitem [{\citenamefont {Moutenet}\ \emph {et~al.}(2018)\citenamefont
  {Moutenet}, \citenamefont {Wu},\ and\ \citenamefont
  {Ferrero}}]{Moutenet2018}%
  \BibitemOpen
  \bibfield  {author} {\bibinfo {author} {\bibfnamefont {A.}~\bibnamefont
  {Moutenet}}, \bibinfo {author} {\bibfnamefont {W.}~\bibnamefont {Wu}}, \ and\
  \bibinfo {author} {\bibfnamefont {M.}~\bibnamefont {Ferrero}},\ }\href
  {\doibase 10.1103/PhysRevB.97.085117} {\bibfield  {journal} {\bibinfo
  {journal} {Phys. Rev. B}\ }\textbf {\bibinfo {volume} {97}},\ \bibinfo
  {pages} {085117} (\bibinfo {year} {2018})}\BibitemShut {NoStop}%
\bibitem [{Note1()}]{Note1}%
  \BibitemOpen
  \bibinfo {note} {We note that $\Sigma ^{\protect \mathrm {I}}_{\alpha z}$
  requires only one correlator, $I_{\alpha z}$, instead of the two needed for
  $\Sigma ^{\protect \mathrm {FG}}_{\alpha z}$. Yet, with the same trick that
  led from Eq.~\protect \textup {\hbox {\mathsurround \z@ \protect \normalfont
  (\ignorespaces \ref {eq:SigmaIG}\unskip \@@italiccorr )}} to \protect \textup
  {\hbox {\mathsurround \z@ \protect \normalfont (\ignorespaces \ref
  {eq:SigmaI}\unskip \@@italiccorr )}}, we can transform Eq.~\protect \textup
  {\hbox {\mathsurround \z@ \protect \normalfont (\ignorespaces \ref
  {eq:SigmaFG}\unskip \@@italiccorr )}} to $\Sigma ^{\protect \mathrm
  {F}}_{\alpha z} = (1+F_{\alpha z})^{-1} F_{\alpha z} (G^0_{\alpha z})^{-1}$.
  This result, too, involves only a single full correlator. However, we
  numerically found $\Sigma ^{\protect \mathrm {F}}_{\alpha z}$ to be less
  accurate than $\Sigma ^{\protect \mathrm {I}}_{\alpha z}$ and hence do not
  discuss it any further.}\BibitemShut {Stop}%
\bibitem [{Note2()}]{Note2}%
  \BibitemOpen
  \bibinfo {note} {One can also use the numerical result obtained for $\protect
  \mathrm {Im}\protect \!\protect \tmspace +\thickmuskip {.2777em}\Sigma
  ^{\protect \mathrm {IFG}}_{\alpha \nu }$ and substitute it on the right of
  Eq.~\protect \textup {\hbox {\mathsurround \z@ \protect \normalfont
  (\ignorespaces \ref {eq:ImSigmaIFG_derivation}\unskip \@@italiccorr )}}
  \protect \textit {instead of} $\protect \mathrm {Im}\protect \!\protect
  \tmspace +\thickmuskip {.2777em}\Sigma ^{\protect \mathrm {FG}}_{\alpha \nu
  }$. This yields a notable improvement at low energies but spoils high-energy
  properties such as the normalization of $\protect \mathrm {Im}\protect
  \!\protect \tmspace +\thickmuskip {.2777em}\Sigma _{\alpha \nu }$. Using
  Eq.~\protect \textup {\hbox {\mathsurround \z@ \protect \normalfont
  (\ignorespaces \ref {eq:ImSigmaIFG_derivation}\unskip \@@italiccorr )}} with
  $\protect \mathrm {Im}\protect \!\protect \tmspace +\thickmuskip
  {.2777em}\Sigma ^{\protect \mathrm {FG}}_{\alpha \nu } \to \protect \mathrm
  {Im}\protect \!\protect \tmspace +\thickmuskip {.2777em}\Sigma ^{\protect
  \mathrm {IFG}}_{\alpha \nu }$ on the right and solving the quadratic equation
  for $\protect \mathrm {Im}\protect \!\protect \tmspace +\thickmuskip
  {.2777em}\Sigma ^{\protect \mathrm {IFG}}_{\alpha \nu }$ did not turn out to
  be helpful.}\BibitemShut {Stop}%
\bibitem [{\citenamefont {{N.~Enenkel \textit{et al.}}}()}]{Enenkel2022}%
  \BibitemOpen
  \bibfield  {author} {\bibinfo {author} {\bibnamefont {{N.~Enenkel \textit{et
  al.}}}},\ }\href@noop {} {\ }\bibinfo {note} {{unpublished}}\BibitemShut
  {NoStop}%
\bibitem [{Note3()}]{Note3}%
  \BibitemOpen
  \bibinfo {note} {In NRG, the value of $\Sigma ^\protect \mathrm {H}_\alpha
  \protect \!=\protect \! F^{(0)}_\alpha $ may slightly differ among $z$
  shifts. Using a $z$-dependent shift $\zeta _\alpha \protect \!=\protect \!
  \Sigma ^\protect \mathrm {H}_\alpha $ makes each term consistently obey
  $\protect \tilde {F}^{(0)}_\alpha \protect \!=\protect \! 0$ and slightly
  improves $\Sigma ^{\protect \mathrm {IFG}}_{\alpha z}$ results.}\BibitemShut
  {Stop}%
\bibitem [{\citenamefont {Luttinger}(1961)}]{Luttinger1961}%
  \BibitemOpen
  \bibfield  {author} {\bibinfo {author} {\bibfnamefont {J.~M.}\ \bibnamefont
  {Luttinger}},\ }\href {\doibase 10.1103/PhysRev.121.942} {\bibfield
  {journal} {\bibinfo  {journal} {Phys. Rev.}\ }\textbf {\bibinfo {volume}
  {121}},\ \bibinfo {pages} {942} (\bibinfo {year} {1961})}\BibitemShut
  {NoStop}%
\bibitem [{\citenamefont
  {Weichselbaum}(2012{\natexlab{a}})}]{Weichselbaum2012a}%
  \BibitemOpen
  \bibfield  {author} {\bibinfo {author} {\bibfnamefont {A.}~\bibnamefont
  {Weichselbaum}},\ }\href {\doibase 10.1016/j.aop.2012.07.009} {\bibfield
  {journal} {\bibinfo  {journal} {Ann. Phys.}\ }\textbf {\bibinfo {volume}
  {327}},\ \bibinfo {pages} {2972} (\bibinfo {year}
  {2012}{\natexlab{a}})}\BibitemShut {NoStop}%
\bibitem [{\citenamefont
  {Weichselbaum}(2012{\natexlab{b}})}]{Weichselbaum2012b}%
  \BibitemOpen
  \bibfield  {author} {\bibinfo {author} {\bibfnamefont {A.}~\bibnamefont
  {Weichselbaum}},\ }\href {\doibase 10.1103/PhysRevB.86.245124} {\bibfield
  {journal} {\bibinfo  {journal} {Phys. Rev. B}\ }\textbf {\bibinfo {volume}
  {86}},\ \bibinfo {pages} {245124} (\bibinfo {year}
  {2012}{\natexlab{b}})}\BibitemShut {NoStop}%
\bibitem [{\citenamefont {Weichselbaum}(2020)}]{Weichselbaum2020}%
  \BibitemOpen
  \bibfield  {author} {\bibinfo {author} {\bibfnamefont {A.}~\bibnamefont
  {Weichselbaum}},\ }\href {\doibase 10.1103/PhysRevResearch.2.023385}
  {\bibfield  {journal} {\bibinfo  {journal} {Phys. Rev. Research}\ }\textbf
  {\bibinfo {volume} {2}},\ \bibinfo {pages} {023385} (\bibinfo {year}
  {2020})}\BibitemShut {NoStop}%
\bibitem [{Note4()}]{Note4}%
  \BibitemOpen
  \bibinfo {note} {Here, we use the well-known formula \cite
  {Haldane1978,Merker2012} for the Kondo temperature $T_{\protect \mathrm {K}}=
  (U\Gamma /2)^{1/2} \protect \qopname \relax o{exp}[ \pi \epsilon _d (\epsilon
  _d+U)/(2U\Gamma ) ]$.}\BibitemShut {Stop}%
\bibitem [{\citenamefont {Haldane}(1978)}]{Haldane1978}%
  \BibitemOpen
  \bibfield  {author} {\bibinfo {author} {\bibfnamefont {F.~D.~M.}\
  \bibnamefont {Haldane}},\ }\href {\doibase 10.1103/PhysRevLett.40.416}
  {\bibfield  {journal} {\bibinfo  {journal} {Phys. Rev. Lett.}\ }\textbf
  {\bibinfo {volume} {40}},\ \bibinfo {pages} {416} (\bibinfo {year}
  {1978})}\BibitemShut {NoStop}%
\bibitem [{\citenamefont {Merker}\ \emph {et~al.}(2012)\citenamefont {Merker},
  \citenamefont {Weichselbaum},\ and\ \citenamefont {Costi}}]{Merker2012}%
  \BibitemOpen
  \bibfield  {author} {\bibinfo {author} {\bibfnamefont {L.}~\bibnamefont
  {Merker}}, \bibinfo {author} {\bibfnamefont {A.}~\bibnamefont
  {Weichselbaum}}, \ and\ \bibinfo {author} {\bibfnamefont {T.~A.}\
  \bibnamefont {Costi}},\ }\href {\doibase 10.1103/PhysRevB.86.075153}
  {\bibfield  {journal} {\bibinfo  {journal} {Phys. Rev. B}\ }\textbf {\bibinfo
  {volume} {86}},\ \bibinfo {pages} {075153} (\bibinfo {year}
  {2012})}\BibitemShut {NoStop}%
\bibitem [{\citenamefont {Mitchell}\ \emph {et~al.}(2014)\citenamefont
  {Mitchell}, \citenamefont {Galpin}, \citenamefont {Wilson-Fletcher},
  \citenamefont {Logan},\ and\ \citenamefont {Bulla}}]{Mitchell2014}%
  \BibitemOpen
  \bibfield  {author} {\bibinfo {author} {\bibfnamefont {A.~K.}\ \bibnamefont
  {Mitchell}}, \bibinfo {author} {\bibfnamefont {M.~R.}\ \bibnamefont
  {Galpin}}, \bibinfo {author} {\bibfnamefont {S.}~\bibnamefont
  {Wilson-Fletcher}}, \bibinfo {author} {\bibfnamefont {D.~E.}\ \bibnamefont
  {Logan}}, \ and\ \bibinfo {author} {\bibfnamefont {R.}~\bibnamefont
  {Bulla}},\ }\href {\doibase 10.1103/PhysRevB.89.121105} {\bibfield  {journal}
  {\bibinfo  {journal} {Phys. Rev. B}\ }\textbf {\bibinfo {volume} {89}},\
  \bibinfo {pages} {121105} (\bibinfo {year} {2014})}\BibitemShut {NoStop}%
\bibitem [{\citenamefont {{Greger}}\ \emph {et~al.}()\citenamefont {{Greger}},
  \citenamefont {{Sekania}},\ and\ \citenamefont {{Kollar}}}]{Greger2013}%
  \BibitemOpen
  \bibfield  {author} {\bibinfo {author} {\bibfnamefont {M.}~\bibnamefont
  {{Greger}}}, \bibinfo {author} {\bibfnamefont {M.}~\bibnamefont {{Sekania}}},
  \ and\ \bibinfo {author} {\bibfnamefont {M.}~\bibnamefont {{Kollar}}},\
  }\href {http://arxiv.org/abs/1312.0100} {\ }\Eprint
  {http://arxiv.org/abs/1312.0100} {arXiv:1312.0100} \BibitemShut {NoStop}%
\bibitem [{\citenamefont {Haule}\ and\ \citenamefont
  {Kotliar}(2009)}]{Haule2009}%
  \BibitemOpen
  \bibfield  {author} {\bibinfo {author} {\bibfnamefont {K.}~\bibnamefont
  {Haule}}\ and\ \bibinfo {author} {\bibfnamefont {G.}~\bibnamefont
  {Kotliar}},\ }\href {http://stacks.iop.org/1367-2630/11/i=2/a=025021}
  {\bibfield  {journal} {\bibinfo  {journal} {New J. Phys.}\ }\textbf {\bibinfo
  {volume} {11}},\ \bibinfo {pages} {025021} (\bibinfo {year}
  {2009})}\BibitemShut {NoStop}%
\bibitem [{\citenamefont {Zhu}\ \emph {et~al.}(2017)\citenamefont {Zhu},
  \citenamefont {Sheng},\ and\ \citenamefont {Zhu}}]{Zhu2017}%
  \BibitemOpen
  \bibfield  {author} {\bibinfo {author} {\bibfnamefont {W.}~\bibnamefont
  {Zhu}}, \bibinfo {author} {\bibfnamefont {D.~N.}\ \bibnamefont {Sheng}}, \
  and\ \bibinfo {author} {\bibfnamefont {J.-X.}\ \bibnamefont {Zhu}},\ }\href
  {\doibase 10.1103/PhysRevB.96.085118} {\bibfield  {journal} {\bibinfo
  {journal} {Phys. Rev. B}\ }\textbf {\bibinfo {volume} {96}},\ \bibinfo
  {pages} {085118} (\bibinfo {year} {2017})}\BibitemShut {NoStop}%
\bibitem [{\citenamefont {Karp}\ \emph {et~al.}(2020)\citenamefont {Karp},
  \citenamefont {Bramberger}, \citenamefont {Grundner}, \citenamefont
  {Schollw\"ock}, \citenamefont {Millis},\ and\ \citenamefont
  {Zingl}}]{Karp2020}%
  \BibitemOpen
  \bibfield  {author} {\bibinfo {author} {\bibfnamefont {J.}~\bibnamefont
  {Karp}}, \bibinfo {author} {\bibfnamefont {M.}~\bibnamefont {Bramberger}},
  \bibinfo {author} {\bibfnamefont {M.}~\bibnamefont {Grundner}}, \bibinfo
  {author} {\bibfnamefont {U.}~\bibnamefont {Schollw\"ock}}, \bibinfo {author}
  {\bibfnamefont {A.~J.}\ \bibnamefont {Millis}}, \ and\ \bibinfo {author}
  {\bibfnamefont {M.}~\bibnamefont {Zingl}},\ }\href {\doibase
  10.1103/PhysRevLett.125.166401} {\bibfield  {journal} {\bibinfo  {journal}
  {Phys. Rev. Lett.}\ }\textbf {\bibinfo {volume} {125}},\ \bibinfo {pages}
  {166401} (\bibinfo {year} {2020})}\BibitemShut {NoStop}%
\bibitem [{\citenamefont {Hafermann}\ \emph {et~al.}(2013)\citenamefont
  {Hafermann}, \citenamefont {Werner},\ and\ \citenamefont
  {Gull}}]{Hafermann2013}%
  \BibitemOpen
  \bibfield  {author} {\bibinfo {author} {\bibfnamefont {H.}~\bibnamefont
  {Hafermann}}, \bibinfo {author} {\bibfnamefont {P.}~\bibnamefont {Werner}}, \
  and\ \bibinfo {author} {\bibfnamefont {E.}~\bibnamefont {Gull}},\ }\href
  {\doibase 10.1016/j.cpc.2012.12.013} {\bibfield  {journal} {\bibinfo
  {journal} {Comput. Phys. Commun.}\ }\textbf {\bibinfo {volume} {184}},\
  \bibinfo {pages} {1280} (\bibinfo {year} {2013})}\BibitemShut {NoStop}%
\bibitem [{\citenamefont {Hafermann}(2014)}]{Hafermann2014}%
  \BibitemOpen
  \bibfield  {author} {\bibinfo {author} {\bibfnamefont {H.}~\bibnamefont
  {Hafermann}},\ }\href {\doibase 10.1103/PhysRevB.89.235128} {\bibfield
  {journal} {\bibinfo  {journal} {Phys. Rev. B}\ }\textbf {\bibinfo {volume}
  {89}},\ \bibinfo {pages} {235128} (\bibinfo {year} {2014})}\BibitemShut
  {NoStop}%
\bibitem [{\citenamefont {Gunacker}\ \emph {et~al.}(2016)\citenamefont
  {Gunacker}, \citenamefont {Wallerberger}, \citenamefont {Ribic},
  \citenamefont {Hausoel}, \citenamefont {Sangiovanni},\ and\ \citenamefont
  {Held}}]{Gunacker2016}%
  \BibitemOpen
  \bibfield  {author} {\bibinfo {author} {\bibfnamefont {P.}~\bibnamefont
  {Gunacker}}, \bibinfo {author} {\bibfnamefont {M.}~\bibnamefont
  {Wallerberger}}, \bibinfo {author} {\bibfnamefont {T.}~\bibnamefont {Ribic}},
  \bibinfo {author} {\bibfnamefont {A.}~\bibnamefont {Hausoel}}, \bibinfo
  {author} {\bibfnamefont {G.}~\bibnamefont {Sangiovanni}}, \ and\ \bibinfo
  {author} {\bibfnamefont {K.}~\bibnamefont {Held}},\ }\href {\doibase
  10.1103/PhysRevB.94.125153} {\bibfield  {journal} {\bibinfo  {journal} {Phys.
  Rev. B}\ }\textbf {\bibinfo {volume} {94}},\ \bibinfo {pages} {125153}
  (\bibinfo {year} {2016})}\BibitemShut {NoStop}%
\end{thebibliography}%

\end{document}